\documentclass[vecphys]{svmult}

\usepackage{makeidx}         
\usepackage{graphicx}        
\usepackage{multicol}        
\usepackage[bottom]{footmisc}
\usepackage{amsmath}
\usepackage{bm}
\usepackage{epsf}
\usepackage{supercite}

\makeindex             

\begin{document}
\sloppy
\title*{Thermodynamics of protein folding from coarse-grained models' perspectives}
\titlerunning{Protein folding from coarse-grained models' perspectives}
\author{Michael Bachmann and Wolfhard Janke}
\institute{Institut f\"ur Theoretische Physik and Centre for Natural Sciences (NTZ), 
Universit\"at Leipzig, Postfach 100\,920, D-04009 Leipzig, Germany\\
\texttt{\{michael.bachmann,wolfhard.janke\}@itp.uni-leipzig.de}}
\maketitle
\section{Introduction}
Proteins are linear chains of amino acids connected by 
covalent peptide bonds. Twenty types of amino acids, mainly differing 
in the molecular structure of their side chains, were identified in 
bioproteins. Since bioproteins typically consist of hundreds to thousands of
amino acids, the number of possible amino acid sequences is extremely large. 
Considering, for example, a chain of only 100 amino acids, the number of
possible sequences (i.e., primary structures) is of order $10^{130}$, but 
this is only one side of the medal. The main importance of 
the proteins lies in their function in a biological organism, and the
function is inseparably connected with the geometrical structure of the protein,
i.e., its folded, native conformation which is usually divided into secondary, 
tertiary, and quarternary substructures~\cite{creighton1,branden1}. The required stability of this native 
state against thermal and other environmental fluctuations rules out most of 
the possible sequences~\cite{dill0}. It is not yet understood, how the relatively small 
number of relevant proteins, e.g., about $10^5$ in the human body, has been 
selected by nature in an evolutionary process~\cite{tang1}. 

The physical interactions responsible for the folding of a protein\index{protein folding} into its 
native structure are in principle known. The complexity of the macromolecule
with up to ten thousands of atoms, however, makes precise 
predictions of the energetically most-favored structure based on ab initio 
quantum-mechanical calculations practically impossible. This is due to the long-range 
overlap of many-body electronic orbitals and the screening by the positively 
charged cores. The problem is indeed still more complex as the natural 
environment of proteins is a polar aqueous solvent. For this reason, classical
models with hundreds of effective parameters (``force fields'') have been developed
in the past decades in order to predict native structures and to study folding
dynamics in computer simulations~\cite{crossrefs1}. Despite the simplifications, these models are
still highly complex and hard to manage even by means of sophisticated 
algorithms and powerful capability computers. 
Furthermore 
it turned out that folding and misfolding depend sensitively on the choice of
the force field parameters with the consequence that predictions of different
established models do frequently not coincide. Another problem is that 
investigations of these models require enormous computational capacities.
For this reason and the fact that folding times in nature range from 
milliseconds to seconds, molecular dynamics simulations (MD) for studying the
deterministic folding dynamics are currently widely useless as the time-scale 
of nano- to microseconds of reliable MD simulations is orders of magnitudes smaller. 

It should be noted,
however, that MD is quite successful in studies of biological short-time 
processes, where the biological function of proteins can be studied.
Fascinating examples, where MD proved to be highly successful, are the penetration 
of water molecules into a cell 
through aquaporin being a membrane protein~\cite{grub1} and the ATP synthase, a process,
where the catalytic subunits of F1, embedded into the membrane F0 proton channel,
partially act as rotating ``molecular motor'' that promotes dehydration of ADP
and P to ATP~\cite{grub2}.
Such studies require that the native folds of the proteins must be known as these are 
used as {\it input}. For considering thermodynamics, Monte Carlo simulations 
of these all-atom models are much more promising, in particular by applying
sophisticated generalized-ensemble algorithms~\cite{genens}. Nonetheless, the enormous efforts
required to obtain trustworthy results with these models strongly limit the 
systematic exposure of the general principles behind protein folding 
processes,
which necessitates comparative studies of an appropriate set of different 
sequences.

In these lecture notes, we therefore follow a different approach and discuss minimalistic, 
coarse-grained protein models. Coarse-graining of models, i.e., increasing 
relevant length scales by reducing the number of microscopic degrees of 
freedom, has proven to be very successful in polymer science.
Although specificity is much more sensitive for proteins, since details (charges, polarity, etc.)
and differences of the amino acid side chains can have strong influences on 
the fold, mesoscopic approaches are also of essential importance
for the basic understanding of conformational transitions affecting the folding process.
It is also the only possible approach for systematic analyses such as the 
evolutionarily significant question
why only a few sequences in nature are 
``designing'', i.e., relevant for selective functions. On the other hand, what
is the reason why proteins prefer a comparative small set of target structures,
i.e., what explains the preference of designing sequences to fold into the 
{\it same} three-dimensional structure? All these questions are widely
still unanswered yet. 

As a first step towards their solution we discuss in the first part simple
hydrophobic-polar (HP) lattice models\index{HP model}, where only the most 
characteristic hydrophobic or polar nature of the 20 naturally occurring 
amino acids is taken into account and the linear chain is modeled by a self-avoiding
walk. Such models allow a comprising analysis
of both, the conformation {\it and} sequence space, e.g., by exactly 
enumerating all combinatorial possibilities. Other important aspects in 
lattice model studies are the identification of lowest-energy conformations
of comparatively long sequences and the characterization of the folding 
thermodynamics.

In the second part we focus on simple AB off-lattice 
models\index{AB model},
where similar to the HP model (for historical reasons) $A$ symbolizes hydrophobic and $B$ polar
regions of the protein, whose conformations are modeled by polymer chains 
governed by bending energy and van der Waals interactions. These models
allow for the analysis of different
mutated sequences with respect to their folding characteristics. Here, the idea
is that the folding transition\index{transition!folding} is a kind of 
pseudophase\index{pseudophase} transition which can
in principle be described by one or a few order-like parameters. Depending on
the sequence the folding process can be highly cooperative (single-exponential),
less cooperative depending on the height of a free-energy barrier 
(two-state folding), or even frustrating due to the existence of different 
barriers in a metastable regime (crystal or glassy phases). These characteristics
known from functional proteins can be recovered in the AB model, which is 
computationally much less demanding than all-atom formulations and thus 
enables throughout theoretical analyses.

Such coarse-grained models enable a broader view on the general problem
of protein folding, but for precise, specific predictions, their applicability
is limited. In analogy to magnetic systems, they are rather comparable with 
the Ising model for ferromagnets or the Edwards-Anderson-Ising model for spin
glasses. It should also be remarked that, due to their nontrivial simplicity, 
coarse-grained models are also a perfect testing ground for newly developed 
algorithms.
\section{Why coarse-graining?}
Functional proteins in a biological organism are typically characterized by a unique 
three-dimensional molecular structure, which makes the protein selective for individual
functions, e.g., in catalytic, enzymatic, and transport processes. In most cases, 
the free-energy landscape is believed to exhibit a rough shape with a large number of
local minima and, for functional proteins, a deep, funnel-like global minimum. This
assumed complexity is the reason, why it is difficult to understand how the random-coil
conformation of covalently bonded amino acids -- the sequence is generated in the
ribosome according to a certain genetic sequence in the DNA -- spontaneously folds
into a well-defined stable ``native'' conformation.
Furthermore, it is expected that there are only a small number of folding paths
from any unfolded conformation to this final fold. 

Protein folding \index{protein folding} follows a strict hierarchy at different length scales. The so-called
primary structure, i.e., the sequence of amino acids in the linear chain is provided 
by the ribosome. Since subsequent amino acids are uniformly linked by a covalent peptide 
bond independent of the geometrical structure of the protein, the typical length scale
of the primary structure is a single amino acid. The next level are secondary structures
like $\alpha$-helices, $\beta$-sheets, and turns. The main reason for the formation of
these structures is backbone hydrogen bonding which typically involves segments
of several subsequent amino acids. Therefore, the scale of secondary structures is 
determined by the typical segment sizes, which are of the order of ten amino acids. Consequently,
secondary-structure formation is the first step in protein folding. This is followed by
the formation of global, single-domain tertiary structures. In fact, this process is 
what renders protein folding special. The main driving force for the folding of a complex
domain, i.e., of up to hundreds of amino acids, is an effective cooperative intrinsic
interaction between many amino acid side-chains and which is strongly influenced by the 
solubility properties (in particular its polarization) of the aqueous solvent the protein 
resides in. Roughly, amino acid side-chains can be classified as polar, hydrophobic, and neutral.
While polar residues favor contact with polar water molecules, hydrophobic acids avoid
contact with water which results in an effective attraction between hydrophobic side-chains.
In consequence, this attractive force leads to a formation of a highly compact hydrophobic core,
which is screened from the solvent by a shell of polar amino acids. For very large proteins,
the final stage in the folding process is the arrangement of several domains in a quarternary
structure.

Thus, the most complex process in protein folding is the formation of tertiary 
hydrophobic-core structures. Although atomic details, e.g., van der Waals volume
exclusion separating side-chains in linear and ring structures, polarizability,
and partial charges, noticeably influence the folding process and the native fold,
it should be possible to understand certain aspects of the folding characteristics,
at least qualitatively, by means of coarse-grained models which
are based on a few effective parameters. In the following, we investigate this question
within the two 
minimalistic HP lattice and AB off-lattice heteropolymer models. 
\section{The hydrophobic-polar (HP) lattice protein model}
The simplest model for a qualitative description of protein folding\index{protein folding} is the lattice
hydrophobic-polar (HP) model~\cite{dill1}\index{HP model}. In this model, the continuous conformational 
space is reduced to discrete regular lattices and conformations of proteins are modeled 
as self-avoiding walks restricted to the lattice. Assuming that the hydrophobic 
interaction is the most essential force towards the native fold,
sequences of HP proteins consist of only two types of monomers (or classes of 
amino acids): Amino acids with high hydrophobicity are treated as 
hydrophobic monomers ($H$), while the class of polar (or hydrophilic) residues
is represented by polar monomers ($P$). In order to achieve the formation 
of a hydrophobic core surrounded by a shell of polar monomers, the interaction
between hydrophobic monomers is attractive and short-range. In the standard formulation of the
model~\cite{dill1}, all other interactions are neglected. Variants of the HP model also take
into account (weaker) interactions between $H$ and $P$ monomers as well as 
between polar monomers~\cite{tang1}. 

Although the HP model is extremely simple, it has been proven that identifying
native conformations is an NP-complete
problem in two and three dimensions~\cite{NPcompl}. Therefore, sophisticated
algorithms were developed to find lowest-energy states for chains of up to 136 
monomers. The methods applied are based on very different algorithms, ranging 
from exact enumeration in two dimensions~\cite{irb1,irb2} and three dimensions
on cuboid (compact) lattices~\cite{tang1,tang2,sbj1,sbj2}, and hydrophobic-core construction
methods~\cite{dill3,dill4} over genetic 
algorithms~\cite{unger1,kras1,cui1,lesh1,jiang1}, Monte Carlo simulations 
with different 
types of move sets~\cite{seno1,rama1,irb3,lee1}, and generalized ensemble 
approaches~\cite{iba1} to Rosenbluth
chain-growth methods~\cite{rosen1} of the {\em 'Go with the Winners'} 
type~\cite{aldous1,grass2,grass3,grass1,hsu1,bj1,bj2}. With some of these algorithms, thermodynamic 
quantities of lattice heteropolymers were studied as well~\cite{sbj1,iba1,grass3,bj1,bj2,naj1}. 
\subsection{The HP model}
A monomer of an HP sequence $\bm{\sigma}=(\sigma_1,\sigma_2,\ldots,\sigma_N)$ 
is characterized by its residual type
($\sigma_i=P$ for polar and $\sigma_i=H$ for hydrophobic residues), 
the position $1\le i\le N$ within the chain of length $N$, and the spatial position ${\bf x}$
to be measured in units of the lattice spacing. A conformation is then symbolized by the
vector of the coordinates of successive monomers, 
${\bf X}=({\bf x}_1,{\bf x}_2,\ldots,{\bf x}_N)$.
We denote by $x_{ij}=|{\bf x}_i-{\bf x}_j|$
the distance between the $i$th and the $j$th monomer.
The bond length between adjacent monomers in the chain is identical with the spacing of the
used regular lattice with coordination number $q$. These covalent bonds are thus
not stretchable.
A monomer and its nonbonded nearest neighbors may form so-called contacts.  
Therefore, the maximum number of contacts of a monomer within the chain is $(q-2)$ and
$(q-1)$ for the monomers at the ends of the chain. To account for the excluded volume,
lattice proteins are self-avoiding, i.e., two monomers cannot occupy the same 
lattice site. The total energy for an HP protein reads 
in energy units $\varepsilon_0$ (we set $\varepsilon_0=1$ in the following)
\begin{equation} 
\label{hpgen} 
E_{\rm HP}=\varepsilon_0\sum\limits_{\langle i, j>i+1 \rangle} C_{ij}U_{\sigma_i\sigma_j},
\end{equation}
where $C_{ij}=(1-\delta_{i+1\,j})\Delta(x_{ij}-1)$ with 
\begin{equation}
\label{defDelta}
\Delta(z)=\left\{ \begin{array}{cl}
1, & \hspace{3mm} z=0,\\
0, & \hspace{3mm} z\neq 0
\end{array}\right. 
\end{equation}
is a symmetric $N\times N$ matrix called {\em contact map} and 
\begin{equation}
\label{intmatrix}
U_{\sigma_i\sigma_j}=\left(\begin{array}{cc}
u_{HH} & u_{HP}\\
u_{HP} & u_{PP} \end{array}\right)
\end{equation}
is the $2\times 2$ interaction matrix. Its elements $u_{\sigma_i\sigma_j}$ correspond to
the energy of $HH$, $HP$, and $PP$ contacts. For labeling purposes we shall adopt
the convention that $\sigma_i=0\,\hat{=}\, P$ and $\sigma_i=1\,\hat{=}\, H$.

In the simplest formulation~\cite{dill1}, only the attractive hydrophobic 
interaction is nonzero, $u^{\rm HP}_{HH}=-1$, while $u^{\rm HP}_{HP}=u^{\rm HP}_{PP}=0$. 
Therefore, $U^{\rm HP}_{\sigma_i\sigma_j}=-\delta_{\sigma_i H}\delta_{\sigma_j H}$. This 
parameterization, which we will traditionally call the {\em HP model}\index{HP model} in the following, 
has been extensively used to identify ground states of HP sequences, some of which are 
believed to show up qualitative properties comparable with realistic proteins whose 20-letter 
sequence was transcribed into the 2-letter code of the HP 
model~\cite{dill3,unger1,shak2,103lat,103toma}. 

This simple form of the standard HP model suffers, however, from the fact that the 
lowest-energy states are usually highly 
degenerate and therefore the number of designing sequences (i.e., sequences with unique 
ground state -- up to the usual translational, rotational, and reflection symmetries) 
is very small, at least on the three-dimensional simple cubic (sc) lattice. Incorporating additional
inter-residue interactions, symmetries are broken, degeneracies are smaller, and 
the number of designing sequences increases~\cite{sbj1,sbj2}.
Based on the Miyazawa-Jernigan matrix~\cite{mj1} of inter-residue contact energies 
between real amino acids, an additional attractive nonzero energy contribution for
contacts between $H$ and $P$ monomers is more realistic~\cite{tang1}. In the following,
we set the elements of the interaction 
matrix~(\ref{intmatrix}) to
$u^{\rm MHP}_{HH}=-1$, $u^{\rm MHP}_{HP}=-1/2.3\approx -0.435$, and $u^{\rm MHP}_{PP}=0$,
corresponding to Ref.~\cite{tang1}. The factor $2.3$ is a result of an analysis 
for the inter-residue energies of contacts
between hydrophobic amino acids and contacts between hydrophobic and polar 
residues~\cite{mj1} which motivated the relation $2 u_{HP}>u_{PP}+u_{HH}$~\cite{tang1}.
In the following we call this variant the {\em MHP model} (mixed HP model). 
\subsection{Exact enumerations for short HP sequences}
The most important advantage of lattice HP-type models compared with other, more complex
protein models is that it allows for comprising analyses of conformation and 
sequence space. This is essential for systematic studies following two main
strategies in understanding protein structure formation: {\em direct} and {\em inverse}
folding. Direct folding is sequence-based, i.e., the amino acid sequence is given and
the global free-energy minimum conformation(s) are sought. In the inverse
folding problem, a target structure is given and the question is for how many sequences
this structure is the global free-energy minimum conformation.

Since it is widely believed that for bioproteins the unique global free-energy minimum conformation 
under physiological conditions (i.e., the native fold) is identical with the conformation of lowest 
total (free) energy,
it is assumed that qualitative folding-related properties of HP lattice protein sequences are 
comparable with realistic proteins if their groundstate is nondegenerate. An HP sequence
with a unique native fold is called {\em designing}. On the other hand, a target structure
which is the native fold of one or more designing sequences, is called a {\em designable conformation}.  
\begin{table}[t]
\caption{\label{tab:desseq} Number of designing sequences $S_N$ (only relevant sequences~\cite{rem1})
in the HP and MHP model on the simple cubic lattice.}
\centerline{
\begin{tabular}{c|*{16}{c}} \hline \hline
$N$ & 4 & 5 & 6 & 7 & 8 & 9 & 10 & 11 & 12 & 13 & 14 & 15 & 16 & 17 & 18 & 19 \\ \hline
$S^{\rm HP}_N$ & 3 & 0 & 0 & 0 & 2 & 0 & 0 & 0 & 2 & 0 & 1 & 1 & 1 & 8 & 29 & 47 \\
$S^{\rm MHP}_N$ & 7 & 0 & 0 & 6 & 13 & 0 & 11 & 8 & 124 & 14 & 66 & 97 & 486 & 2196 & 9491 & 4885 \\ \hline \hline
\end{tabular}
}
\end{table}
\begin{table}[t]
\caption{\label{tab:design} Number of designable conformations $D_N$ (without conformations 
trivially symmetric by translations, rotations, and reflections) in the HP and MHP model
on the simple cubic lattice.}
\centerline{
\begin{tabular}{c|*{16}{c}} \hline \hline
$N$ & 4 & 5 & 6 & 7 & 8 & 9 & 10 & 11 & 12 & 13 & 14 & 15 & 16 & 17 & 18 & 19 \\ \hline
$D^{\rm HP}_N$ & 1 & 0 & 0 & 0 & 2 & 0 & 0 & 0 & 2 & 0 & 1 & 1 & 1 & 8 & 28 & 42 \\ 
$D^{\rm MHP}_N$ & 1 & 0 & 0 & 2 & 2 & 0 & 5 & 6 & 30 & 8 & 31 & 58 & 258 & 708 & 1447 & 1623 \\ \hline \hline
\end{tabular}
}
\end{table}

In Table~\ref{tab:desseq} we list for all chain lengths $N=4,\ldots,19$ the total numbers $S_N$ 
of relevant designing sequences~\cite{rem1} in the HP and the MHP model. These results were obtained
by exhaustive exact enumerations of the complete conformation and sequence spaces of chains with up to
19 monomers on the sc lattice~\cite{sbj1}. Note that there are for a 19-mer more than $5\times 10^5$ 
HP sequences and about $2\times 10^{12}$ self-avoiding conformations on the sc lattice, which in total
allows naively more than $10^{18}$ possible combinations. In order to achieve this, an 
efficient parallel implementation based on contact sets~\cite{irb2,domany1} together with
symmetry considerations had to be used~\cite{sbj2}.
As already mentioned, the number of designing sequences is rather small in the standard HP model, whereas
the additional $HP$ attraction in the MHP model dissolves degeneracies which consequently entails
a noticeably larger number of sequences with a unique ground-state conformation.
\begin{figure}[t]
\centerline {
\epsfxsize=11cm \epsfbox {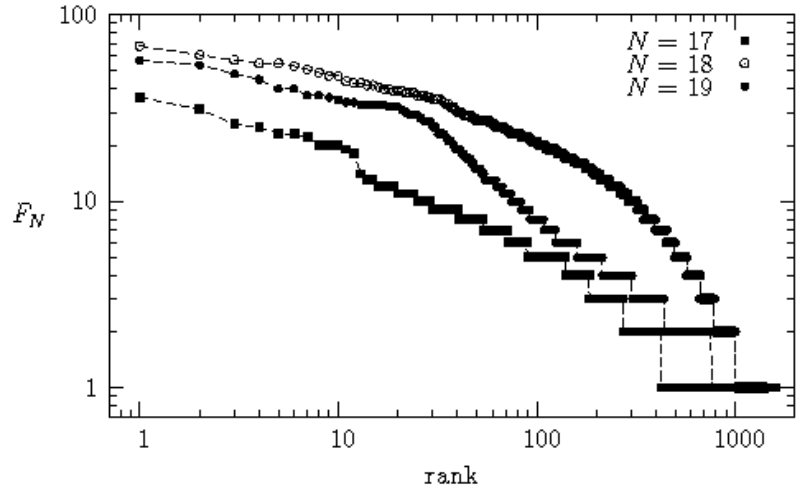}
}
\caption{\label{fig:design} 
Designability $F_N$ of native conformations 
in the MHP model for $N=17$, $18$, and $19$. The abscissa is the rank obtained 
by ordering all designable conformations according to their designability.
}
\end{figure}

In Table~\ref{tab:design} we list for both models 
the number of {\em different} native conformations $D_N$ on the sc lattice. 
Interestingly, this number is usually much smaller than the number of designing sequences
in Table~\ref{tab:desseq}, 
i.e., several designing sequences share the same ground-state conformation. The number of
designing sequences that fold into a certain given target conformation 
${\bf X}^{(0)}$ (or conformations being trivially symmetric to this by translations, 
rotations, and reflections) is called {\em designability}~\cite{tang4}:
\begin{equation}
\label{design}
F_N({\bf X}^{(0)})=\sum\limits_{\bm{\sigma}\in\mathbf{S}_N}
\Delta\left({\bf X}_{\rm gs}(\bm{\sigma})-{\bf X}^{(0)}\right),
\end{equation}
where ${\bf X}_{\rm gs}(\bm{\sigma})$ is the native (ground-state) 
conformation of a designing sequence
$\bm{\sigma}$ in the set of all designing sequences $\mathbf{S}_N$ of length $N$. 
The function $\Delta({\bf Z})$ is the generalization of Eq.~(\ref{defDelta})
to $3N$-dimensional vectors. It is unity for ${\bf Z}={\bf 0}$ and zero otherwise.
\begin{figure}[t]
\centerline{
\hfill\parbox{4.5cm}{
\epsfxsize =4.3cm \epsfbox {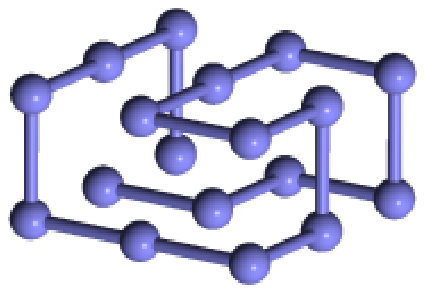}}
\hfill
\parbox{4.5cm}{
\epsfxsize =4.3cm \epsfbox {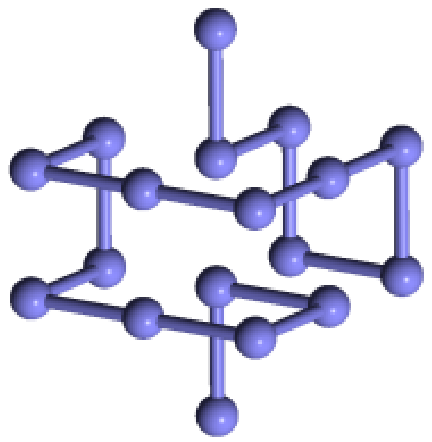}}
\hfill}
\caption{\label{fig:highmax} 
Structure ($N=18$) with the highest designability of all native 
conformations (left) and most compact structure with minimal radius of gyration (right).
}
\end{figure}

The designability is plotted in Fig.~\ref{fig:design} for all native
conformations that HP proteins with $N=17$, $18$, and $19$ monomers can form in the MHP
model. In this figure, 
the abscissa is the rank of the conformations, ordered according to their 
designability. The conformation with the lowest rank is therefore the most designable structure
and we see that most of the designing sequences fold into a few number of 
highly designable conformations, while only a small number of designing sequences 
possesses a native conformation with low designability (note that the plot is logarithmic). 
Similar results were found, for example in Ref.~\cite{tang3}, where the designability of
compact conformations on cuboid lattices was investigated in detail. The left picture
in Fig.~\ref{fig:highmax} shows the conformation with the lowest rank (or highest designability) 
with $N=18$ monomers. Note that this is not the most compact structure, i.e., the conformation
with minimal gyration radius, which is shown for $N=18$ in Fig.~\ref{fig:highmax} (right). 
\subsection{Chain-growth methods for long HP sequences}
\label{ssec:cgmeth}
Combined exact enumeration studies of conformation {\em and} sequence space for lattice peptides 
noticeably longer than 19 monomers are currently computationally out of reach which is due to
the exponential growth of the state space. Therefore, for longer sequences, primarily the direct
folding problem is studied using computer simulation methods: 
Low-lying energetic conformations and thermodynamic properties
governing the folding kinetics are identified and analyzed for a given HP sequence.

Computer simulations of lattice peptides, which are modeled as self-avoiding
walks on the underlying lattice, are  demanding. The reason is that the native fold,
i.e., the ground-state or lowest-energy conformation, plays an essential role in
protein science and that it is, in the discrete lattice representation, non- or
low-degenerate. Monte Carlo simulations with move sets consisting of semilocal
conformational updates like end flips, corner flips, 
and ``crank shafts''~\cite{seno1,rama1,irb3,lee1}, as well
as nonlocal pivot updates~\cite{madsok1}, are inefficient in sampling the dominating dense
conformations in the low-temperature region. It turned out that a different method,
Rosenbluth chain growth~\cite{rosen1}\index{chain-growth!Rosenbluth} combined with a 'Go with the winners' 
strategy~\cite{aldous1}, is much more efficient in sampling highly dense conformations. 
\subsubsection{Pruned-enriched Rosenbluth chain-growth method (PERM)}
\index{PERM}\index{chain-growth!pruned-enriched}
In naive chain-growth methods based on simple sampling, a polymer grows by attaching the 
$n$th monomer at a randomly chosen nearest-neighbor 
site of the $(n-1)$th monomer. The growth is stopped, if the total length $N$
of the chain is reached or the randomly selected continuation of the chain is already
occupied. In both cases, the next chain is started to grow
from the first monomer. This simple chain growth is not yet very efficient, since 
the number of discarded chains grows exponentially with the chain length.

\begin{figure}[t]
\centerline{\hfill
\parbox{4cm}{
\centerline{\epsfxsize =3.5cm \epsfbox {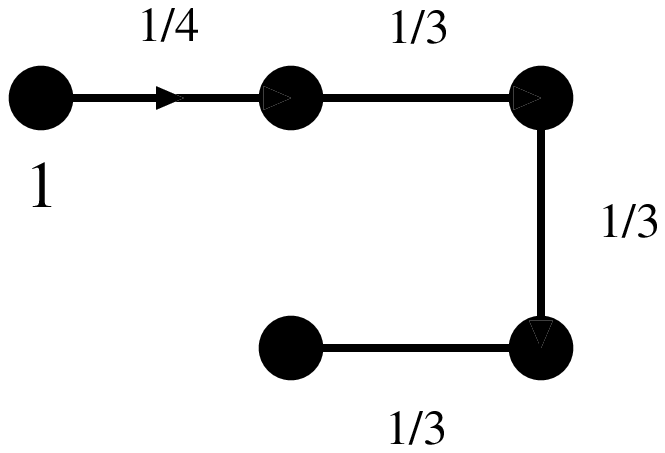}}}
\hfill
\parbox{4cm}{
\centerline{\epsfxsize =3.5cm \epsfbox {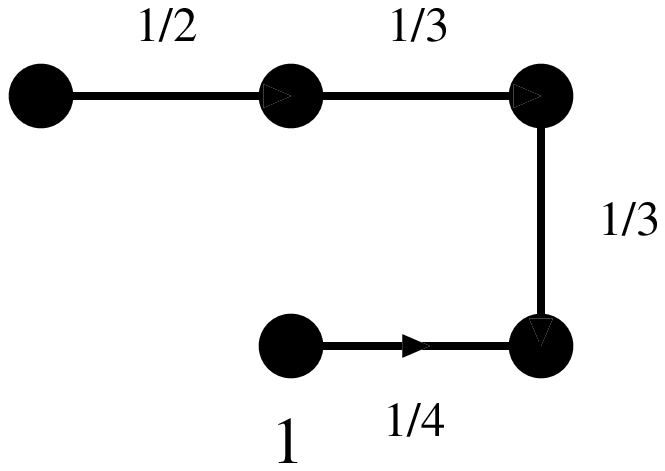}}}
\hfill}
\caption{\label{fig:rosen} Square-lattice example for the bias implied by Rosenbluth sampling. Both 
walks shown are grown from the monomer labeled ``1''. Although the shapes are identical, they are created 
with different probabilities (left: $p=1/108$, right: $p=1/72$).
}
\end{figure}
The performance can be improved with the Rosenbluth chain growth method~\cite{rosen1}, where
first the free next neighbors of the 
$(n-1)$th monomer are determined and then the new monomer is placed to one of 
the unoccupied sites. 
Since the probability for the next monomer
to be set varies with the number of free neighbors, this implies a bias given by
\begin{equation}
p_n= \left(\prod_{l=2}^n m_l\right)^{-1},
\end{equation}
where $m_l$ is the number of free neighbors to place the $l$th monomer. 
The bias 
is corrected by assigning a Rosenbluth weight factor $W_n^{\rm R}= p_n^{-1}$ 
to each chain that has been generated by this procedure. An illustrative example 
for the bias in the Rosenbluth chain-growth method
is shown in Fig.~\ref{fig:rosen}. The two depicted linear chains are grown on a square lattice
from both ends (labeled by ``1''). According to Rosenbluth sampling, the chain is continued if
the number of free neighbor sites is $m\ge 1$. Since the number of free nearest-neighbor places varies,
different probabilities for the continuation of the chain occur. Since both conformations are 
identical, the probability of creation should be the same. This requires the introduction of the
correction weights.   
Although this biased growth is more efficient than simple sampling, this method suffers 
from attrition too: If all nearest neighbors are 
occupied, i.e., the chain was running into a ``dead end'' (attrition point), the complete chain
has to be discarded and the growth process has to be started anew.  

Combining the Rosenbluth chain growth method with population control, however, as is done 
in PERM (Pruned-Enriched Rosenbluth Method)~\cite{grass2,grass3,grass1}\index{PERM}, leads to
a further considerable improvement of the efficiency by increasing the number of successfully
generated chains. This method renders particularly useful for studying the $\Theta$ point 
of polymers, since then the Rosenbluth weights of the statistically relevant chains 
approximately cancel against their Boltzmann probability. The (a-thermal) Rosenbluth 
weight factor $W_n^{\rm R}$ is therefore replaced by 
\begin{equation}
\label{PERMweight}
W_n^{\rm PERM}=\prod\limits_{l=2}^n m_l e^{-(E_l-E_{l-1})/k_BT},\quad 
 2\le n\le N \quad (E_1=0, \quad W_1^{\rm PERM}=1), \nonumber
\end{equation} 
where $T$ is the temperature and $E_l$ is the energy of the partial chain 
${\bf X}_l=({\bf x}_1,\ldots ,{\bf x}_l)$ created with Rosenbluth chain growth.
In PERM, population control works as follows. If a chain has reached length $n$,
its weight $W_n^{\rm PERM}$ is calculated and compared with suitably chosen
upper and lower threshold values, $W_n^>$ and $W_n^<$, respectively. For 
$W_n^{\rm PERM} > W_n^>$, {\em identical} copies are created which grow then
independently. The weight is equally divided among them. If $W_n^{\rm PERM} < W_n^<$,
the chain is pruned with some probability, say 1/2, and in case of survival, its 
weight is doubled. For a value of the weight lying between the thresholds, the
chain is simply continued without enriching or pruning the sample. The  
upper and lower thresholds $W_n^>$ and $W_n^<$ are empirically parameterized.
Although their values do not influence the validity of the method, a careful
choice can drastically improve the efficiency of the method (the ``worst'' case
is $W_n^>=\infty$ and $W_n^<=0$, in which case PERM is simply identical with
Rosenbluth sampling). An
efficient way of parameterization is dynamical adaption of the 
values~\cite{grass2,grass3,grass1,hsu1,bj1,bj2} with respect to the actual number 
of generated chains $c_n$ with length $n$ and their estimated partition sum
\begin{equation}
\label{PERMpartsum}
Z_n=\frac{1}{c_1}\sum\limits_t W_n^{\rm PERM}(t),
\end{equation} 
where $c_1$ is the number of growth starts (also called ``tours'') 
and $t$ counts the generated conformations
with $n$ monomers. Useful choices of the threshold values are 
\begin{equation}
\label{PERMthresh}
W_n^>=C_1\, Z_n\frac{c_n^2}{c_1^3},\quad W_n^<=C_2\, W_n^>,
\end{equation}
where $C_1,C_2\le 1$ are constants. For the first tour, $W_n^>=\infty$ and $W_n^<=0$,
i.e., no pruning and enriching.

\begin{figure}[t]
\centerline {
\epsfxsize=11cm \epsfbox {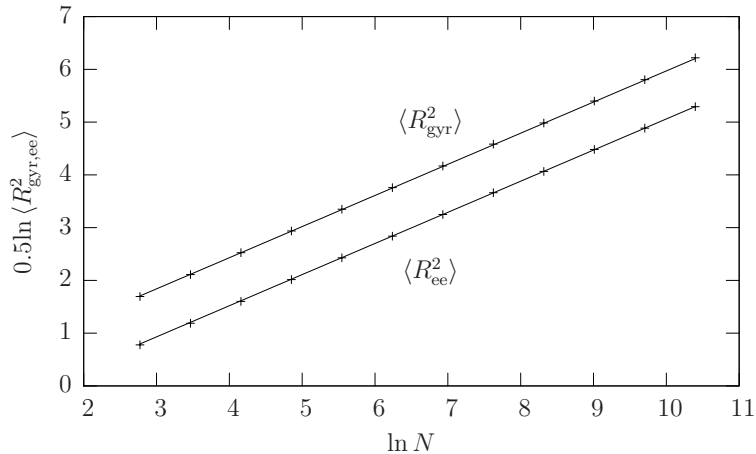}
}
\caption{\label{fig:saw} 
Scaling of mean square radius of gyration $\langle R^2_{\rm gyr}\rangle$ and end-to-end distance 
$\langle R^2_{\rm ee}\rangle$ for self-avoiding walks. Data points refer to results from PERM
runs for $N=16,32,\ldots,32768$ steps. Lines manifest the respective power-law behaviors.
}
\end{figure}
In the recently developed new variants nPERMss and nPERMis~\cite{hsu1},
the number of copies is not constant and depends on the ratio of the weight 
$W_n^{\rm PERM}$ compared to the upper threshold value $W_n^>$
and the copies are necessarily chosen to be different. The method of selecting
the copies is based on simple sampling (ss) in \mbox{nPERMss} and a kind of 
importance sampling (is) in \mbox{nPERMis}. This proves quite
useful in producing highly compact polymers and therefore these new methods
are very powerful in determining lowest-energy states of lattice proteins. 

Results of a simple application of PERM to self-avoiding walks on a simple-cubic lattice
are plotted in Fig.~\ref{fig:saw},
where the scaling behavior $\langle R^2_{\rm gyr,ee}\rangle\sim N^{2\nu}$
of the mean square radius of gyration $\langle R^2_{\rm gyr}\rangle$
and end-to-end distance $\langle R^2_{\rm ee}\rangle$ with the number of steps $N$ is shown.
Data were obtained for chains of $N=16,32,\ldots,32768$ steps. For both quantities, the slope of the lines
in the logarithmic plot is $\nu= 0.59$, which is close to the precisely known critical exponent 
$\nu=0.588\ldots$~\cite{guillou1}.
\subsubsection{Multicanonical chain-growth algorithm}
\index{chain-growth!multicanonical}
The efficiency of PERM depends on the simulation temperature. Therefore, a precise 
estimation of the density of states requires separate simulations at different 
temperatures. Then, the density of states can be constructed by means of the
multiple-histogram reweighting method~\cite{ferr1}. Although being a powerful method,
it is difficult to keep track of the statistical errors involved in the individual
histograms obtained in the simulations.

An alternative approach, in which the density of states $g(E)$ is obtained within a {\em single}
simulation without the necessity of a subsequent multi-histogram reweighting, is
the combination of PERM with multicanonical sampling, the so-called
multicanonical chain-growth method~\cite{bj1,bj2}. 

The general idea of
multicanonical sampling~\cite{muca1,muca2}\index{multicanonical sampling} is to simulate the thermodynamic behavior of
the system in a generalized (multicanonical) ensemble, where the energetic macrostates
are distributed uniformly, $p_{\rm muca}(E)={\rm const}$, which implies the introduction of
multicanonical weight factors $W_{\rm muca}(E)$. In typical multicanonical Monte Carlo simulations, the 
dynamics is therefore governed by a random walk in energy space. Hence, the sampling 
of entropically rare events is, in principle, as frequent as the sampling of highly 
degenerate energetic states. The acceptance probability for a 
new system configuration ${\bf X'}$ with energy $E'$ is 
$w_{\rm muca}({\bf X}\to {\bf X'})=\min[1, \exp\{S(E({\bf X'}))-S(E({\bf X}))\}]$,
where $S(E({\bf X}))=-{\rm ln}\,W_{\rm muca}(E({\bf X})$ is the 
microcanonical entropy. The canonical energy distribution 
$p_{\rm can}(E)\sim g(E)\exp(-E/k_BT)$ for a 
given temperature $T$ is related with the multicanonical histogram via
\begin{equation}
\label{eq:muca}
p_{\rm can}(E)\sim W_{\rm muca}^{-1}(E) p_{\rm muca}(E) e^{-E/k_BT},
\end{equation}
which implies that the multicanonical weights are proportional to the inverse density
of states, $W_{\rm muca}(E)\sim g^{-1}(E)$. Since $g(E)$ is unknown, the determination of 
the weights $W_{\rm muca}(E)$ is not straightforward and must be performed in the first stage
of the simulation in an iterative procedure~\cite{muca2}. 

The multicanonical extension of PERM requires two main changes compared to standard PERM. 
Firstly, the expression~(\ref{PERMweight}) for the 
weight factor is replaced by
\begin{equation}
\label{eq:mweight}
W_n^{\rm MPERM}(E_n)=W_{{\rm muca},n}(E_n)\prod\limits_{l=2}^n m_l,\quad
W_{{\rm muca},1}= 1,
\end{equation}
where, according to multicanonical sampling, the multicanonical weight of the chain of current length $n$ is related to 
the appropriate inverse density of states, $W_{{\rm muca},n}(E)\sim g^{-1}_n(E)$.
Note that the possibility to rewrite Eq.~(\ref{eq:mweight}) in the recursive, factorized form
\begin{equation}
\label{eq:fmweight}
W_n^{\rm MPERM}(E_n)=\prod\limits_{l=2}^n m_l \frac{W_{{\rm muca},l}(E_l)}{W_{{\rm muca},l-1}(E_{l-1})}
=W_{n-1}^{\rm MPERM} m_n\frac{W_{{\rm muca},n}(E_n)}{W_{{\rm muca},n-1}(E_{n-1})}
\end{equation}
is mainly responsible for the efficiency of this method as it ensures that rare-event (flat-histogram)
sampling is performed in {\em all} intermediate steps of the growth process. This means that
for a chain of length $N$ all energy histograms are ``flat'', $H_n(E)\approx {\rm const.}$ with $n\le N$.
The pruning-enriching scheme of PERM is completely carried over and remains unchanged with the 
exception that the thresholds~(\ref{PERMthresh}) are re-expressed as
\begin{equation}
\label{eq:MPERMthresh}
W_n^>=C_1\, Z_n^{\rm MPERM}\frac{c_n^2}{c_1^3},\quad W_n^<=C_2\, W_n^>,
\end{equation}
i.e., in terms of the partition sum of the multicanonical ensemble, 
$Z_n^{\rm MPERM}=\sum\limits_t W_n^{\rm MPERM}(t)/c_1$.

\begin{figure}[t]
\centerline {
\epsfxsize=11cm \epsfbox {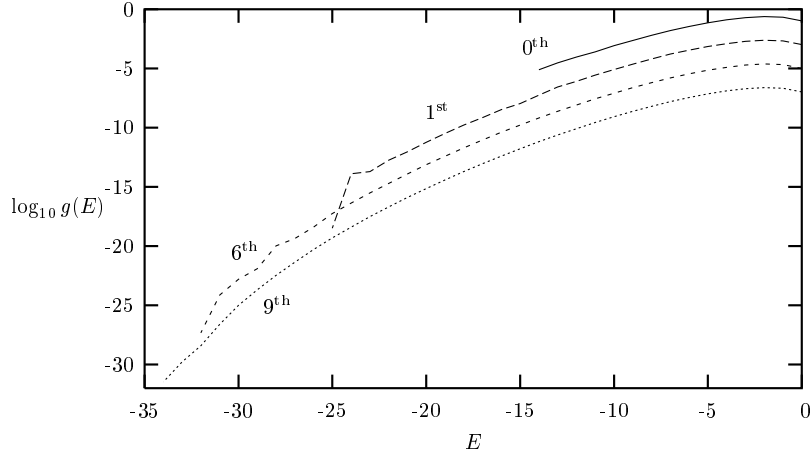}
}
\caption{\label{fig:42iter} 
Estimates for the density of states $g^{(i)}(E)$ for an exemplified heteropolymer with 42 monomers
after several recursion levels.  Since the curves would fall on top of each
other, we have added, for better distinction, a suitable offset to
the curves of the $1$th, $6$th, and $9$th run. The estimate of the $0$th run is normalized
to unity.
}
\end{figure}
The second difference compared with the original PERM is the estimation of the multicanonical weights,
as the densities of
states $g_n(E)$, $n\le N$, are unknown in the beginning of the simulation. Therefore, the 
multicanonical weight factors $W_{{\rm muca},n}(E)$ must be determined iteratively
for all stages $n\le N$ of the growth process~\cite{bj2}. The initial choice for the 
multicanonical weights is typically $W^{(0)}_{{\rm muca},n}(E)=1\, \forall n,E$, making
the zeroth recursion a pure PERM run at infinite temperature. The energy histograms are initialized
with $H^{(0)}_n(E)=0$. Performing the multicanonical chain growth according to the method described
above, the histograms are accumulated by summing up the weights~(\ref{eq:fmweight}) of successively
generated chains:
\begin{equation}
\label{eq:histzero}
H^{(0)}_n(E)=\frac{1}{c_1}\sum\limits_{t}W_{n,t}^{\rm MPERM}\,\delta_{E_t\, E},
\end{equation}
where $t$ labels the chain reaching length $n$ in the growth process. Since this histogram is a first
estimate for the density of states, the multicanonical weights for the following iteration are set to
$W^{(1)}_{{\rm muca},n}(E)=1/H^{(0)}_n(E)$. Before starting the new recursion, 
$Z_n^{\rm MPERM}$, $c_n$, $W_n^<$ are reset to zero, and $W_n^>$ to infinity (i.e.\ to the upper limit
of the data type used to store this quantity). The iterative procedure is repeated until the 
weights $W^{(i)}_{{\rm muca},n}(E)=W^{(i-1)}_{{\rm muca},n}(E)/H^{(i-1)}_n(E)$ are stabilized. 
In a long final production run $i=I$, the densities of states are then determined from
\begin{equation}
\label{eq:densMPERM}
g_n^{(I)}(E)=\frac{H_n^{(I)}(E)}{W_{{\rm muca},n}^{(I)}(E)},\quad n\le N.
\end{equation} 
For practical applications of this algorithm, in particular for studies of heteropolymers, it
is more favorable to replace the original pruning-enrichment core, i.e., PERM~\cite{grass2}, by the modern,
improved variants nPERMss or \mbox{nPERMis}~\cite{hsu1}. The combination of this more efficient 
chain-growth strategy with multicanonical sampling is straightforward. The details are explained
in Refs.~\cite{bj1,bj2}.
In Fig.~\ref{fig:42iter}, estimates for $g^{(i)}(E)$ after the iterations $i=0,1,6$, and $9$ are 
shown for an exemplified heteropolymer with 42 monomers, whose thermodynamic properties will be 
discussed in more detail in Section~\ref{ssec:hpthermo}. The zeroth run is a pure PERM
estimate with a reliable precision over 5 orders of magnitude. As the simulations were effectively
performed in the purely entropic regime at infinite temperature (i.e., $\beta = 0$), 
low-energy states are rarely sampled. In this case, chain growth is governed only 
by the weights~(\ref{PERMweight}) which are only products of free nearest-neighbor 
sites and therefore identical with the Rosenbluth weights for self-avoiding walks. 
The model-dependent energetic influence on the growth is thus irrelevant. The efficiency is improved 
in the successive recursions, where the multicanonical weights~(\ref{eq:fmweight}) are
gradually refined and allow for a sampling of larger regions of the energy space. After
only 10 recursions, the estimate for $g(E)$ covers the whole accessible energy space (including
the ground state) and ranges over 25 orders of magnitude. 

It is worth noticing that the obtained density of states is {\em absolute}, i.e.,
estimates for the degeneracies of energetic states, in particular for the important ground state, can
directly be read off. Furthermore, the partition function is also absolutely estimated via
$Z=\sum_E g(E)\exp(-E/k_BT)$. The reason is that these chain-growth methods perform a ``biased'' 
simple-sampling instead of importance sampling as is used in most Monte Carlo methods. With importance
sampling, it is usually not possible to obtain an absolute estimate for $g(E)$.

The probability for energetic states in a canonical ensemble at temperature $T$ 
is obtained from the density of states by Boltzmann reweighting
via $p_{\rm can}(T)=g(E)\exp(-E/k_BT)/Z$. Thus, statistical expectation values of energetic 
observables $O(E)$ are simply given by $\langle O\rangle=\sum_E O(E)p_{\rm can}(E)$. Thermal fluctuations
of these quantities, e.g., defined 
by $d\langle O\rangle/dT=(\langle O^2\rangle-\langle O\rangle^2)/k_BT^2$, are of particular interest 
for identifying temperature regions of thermodynamic activity. A very convenient measure 
for quantifying the cooperative behavior of a complex system is, e.g., 
the specific heat $C_V=(\langle E^2\rangle-\langle E\rangle^2)/k_BT^2$.  
\subsubsection{Decoupling energy scales: An instructive example}
For systems, where different energy scales decouple, the density of states $g(E)$ as a distribution 
of states with given {\em total} 
energy $E$ is not the most useful quantity. As an important example, we consider the adsorption of a polymer 
to a substrate. In simple lattice models, only the number of intrinsic nearest-neighbor 
contacts between nonadjacent monomers, $n_m$, and the number of nearest-neighbor contacts of the polymer
with the substrate, $n_s$, are counted. For the discussion of conformational transitions\index{transition!conformational} 
in the adsorption
process later on, it is quite useful to rate intrinsic and binding forces against each other and therefore
it is useful to introduce different energy scales $\varepsilon_m$ and $\varepsilon_s$ 
corresponding to the contact numbers $n_m$ and $n_s$, respectively. A minimalistic model
could then, for example, be defined by~\cite{vrbova}
\begin{equation}
\label{eq:adsmodel}
E(n_m,n_s)=-\varepsilon_m n_m-\varepsilon_s n_s\equiv -\varepsilon_0(sn_m+n_s),
\end{equation}
where the ratio $s=\varepsilon_m/\varepsilon_s$ can be considered as kind of reciprocal solvent parameter 
(the larger $s$, the worse the quality of the solvent). The overall energy scale is simply
$\varepsilon_0\equiv \varepsilon_s$. Since the total energy $E$ of the system depends
on $s$, it would be necessary to fix its value in the previously described multicanonical chain-growth 
variant. Instead of determining the density of states $g(E)$, it would be more favorable to calculate
the {\em contact density} $g(n_m,n_s)$ which is independent of $s$. Knowing the contact density, the
canonical probability for a system conformation with $n_m$ monomer-monomer and $n_s$ monomer-substrate
contacts is given by $p_{T,s}(n_m,n_s)=g(n_m,n_s)\exp[-E(n_m,n_s)/k_BT]/Z_{T,s}$, where temperature
$T$ and solubility $s$ are considered as fixed parameters. The statistical average of a 
quantity $O(n_m,n_s)$ is then obtained as $\langle O(n_m,n_s)\rangle=\sum_{n_m,n_s} O(n_m,n_s)p_{T,s}(n_m,n_s)$.
For the discussion of the conformational-phase diagram of the hybrid polymer-substrate
system in solvent, it is useful to consider the dependence of fluctuations on temperature {\em and} solubility. 
As an example, the specific heat can be expressed as
\begin{equation}
\label{eq:adssh}
C_V(T, s)=k_B\left(\frac{\varepsilon_0}{k_BT}\right)^2\ (s\ 1)
\left(\begin{array}{cc}
\langle n_s^2\rangle_c & \langle n_s n_m\rangle_c\\
\langle n_s n_m\rangle_c & \langle n_m^2\rangle_c
\end{array}\right)
\left(\begin{array}{c}
s \\ 1
\end{array}\right),
\end{equation}
where $\langle xy\rangle_c=\langle xy\rangle-\langle x\rangle\langle y\rangle$ ($x,y=n_m,n_s$)
are the variances and covariances of the contact numbers.
Note that the knowledge of $g(n_m,n_s)$ enables reweighting of the specific heat to any pair
of parameters $T$ and $s$.
\subsubsection{Contact density chain-growth method}
\index{chain-growth!contact density}
The determination of the contact density $g(n_m,n_s)$ follows similar lines as the multicanonical chain-growth
method for the estimation of the density of states. In fact, the only change in the algorithm 
described in the previous section is that the weights $W_n^{\rm MPERM}(E_n)$ defined in 
Eq.~(\ref{eq:fmweight}) are replaced by
\begin{equation}
\label{eq:mucoweight}
W_n^{\rm CDPERM}(n_m^{(n)},n_s^{(n)}) = \prod\limits_{l=2}^n m_l 
\frac{W_{{\rm cd},l}(n_m^{(l)},n_s^{(l)})}{W_{{\rm cd},l-1}(n_m^{(l-1)},n_s^{(l-1)})},
\end{equation}
where the multi-contact weights $W_{{\rm cd},l}(n_m^{(l)},n_s^{(l)})\sim 1/g(n_m^{(l)},n_s^{(l)})$ have again
to be determined recursively.

The extension of this method incorporating more than two system parameters is straightforward, but 
the efficiency of flattening the high-dimensional histograms at all levels of the growth process decreases,
whereas the storage requirements for these fields rapidly increase. 
\subsection{Bulk behavior of HP lattice proteins}
\label{ssec:hpthermo}
Before embarking into the discussion of hybrid peptide-substrate systems, we investigate
first the bulk behavior of HP peptides.
In the folding process from a random-coil conformation towards the native fold, the protein
experiences in many cases conformational transitions\index{transition!conformational}. These transitions typically require 
passing or circumventing of barriers in the free-energy landscape, which slows down the folding 
dynamics. Similar to thermodynamic phase transitions, conformational transitions can be
identified by noticeable changes in the behavior of fluctuating quantities. Peaks
and ``shoulders'' in specific-heat curves are, for example, typical signals for
cooperative activity, because in the vicinity of the peak temperatures entropic changes 
separate qualitatively different classes of conformations (e.g., random coils and globular
shapes). Since peptides are always of finite length due to their well-defined
amino acid sequence, conformational transitions are not phase transitions in the strict
thermodynamic sense. In consequence, fluctuations of different thermodynamic quantities
typically do not exhibit the same peak structure, i.e., there is no ``data collapse'' which
would allow the definition of a uniform transition temperature, where the phases are
uniquely separated. Since for peptides different fluctuating quantities predict different
transition temperatures, it is only possible to identify a temperature interval of
conformational activity. This makes a precise quantitative analysis and a qualitative 
classification of such transitions difficult~\cite{bj1,bj2}. 

\begin{figure}[t]
\centerline {
\epsfxsize=11cm \epsfbox {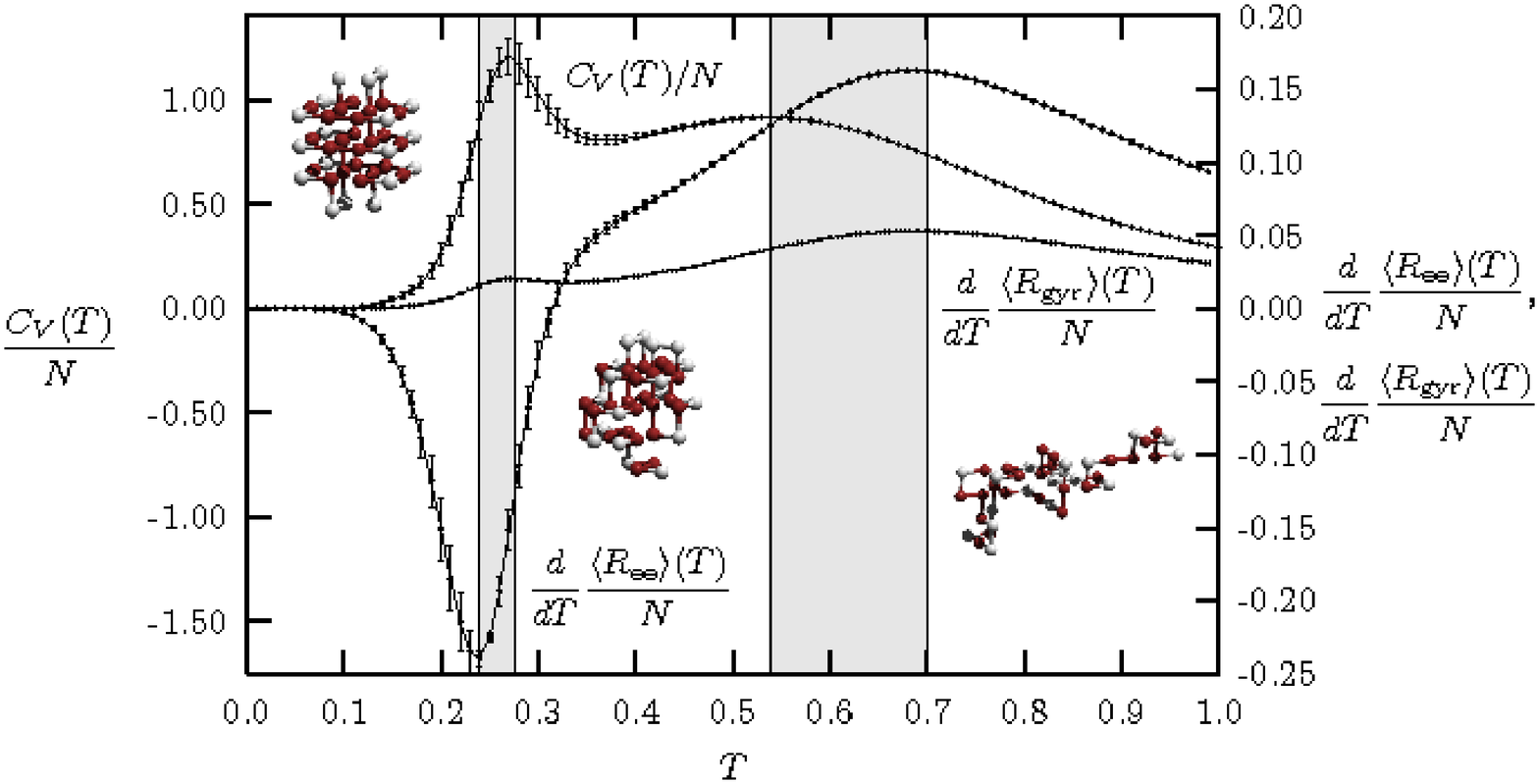}
}
\caption{\label{fig:42fl} 
Specific heat $C_V$ and respective fluctuations of gyration radius and end-to-end distance,
$d\langle R_{\rm gyr}\rangle/dT$ and $d\langle R_{\rm ee}\rangle/dT$, as functions of temperature
for the 42-mer. 
}
\end{figure}
In the following, we discuss ground-state properties and 
thermodynamics for exemplified HP sequences. These results were obtained by employing the 
aforementioned multicanonical chain-growth method~\cite{bj1,bj2}\index{chain-growth!multicanonical} 
for the standard version of the HP 
model~\cite{dill1}.
An interesting example is the 42-monomer HP sequence representing the parallel $\beta$-helix 
protein {\em pectate lyase C}~\cite{yoder1}, which reads
PH$_2$PHP\-H$_2$\-P\-H\-P\-H\-P$_2$\-H$_3$\-P\-H\-P\-H$_2$\-P\-H\-P\-%
H$_3\-$P$_2$\-H\-P\-H\-P\-H$_2$\-P\-HPH$_2$P~\cite{dill3}. Although it is not believed that 
specific protein properties such as the folding behavior and thermodynamics are conserved in
a one-to-one transcription of an amino acid sequence into the hydrophobic-polar two-letter
code, this example shows surprising coincidences of the real protein and the model, 
as the (low-degenerate) ground-state conformations in the HP model also exhibit two 
parallel helical segments. More interesting is,
however, that the ground-state degeneracy is only $g_0=4$ without trivial rotational 
symmetries~\cite{dill3}. With multicanonical chain-growth 
simulations~\cite{bj1}, the ground-state degeneracy was 
precisely estimated as $g_0=3.9\pm 0.4$~\cite{bj2}. In this simulation, 
10 recursions were performed and in the production run, about $5\times 10^7$ chains
entered into statistics. 

The low ground-state degeneracy is indeed remarkable, as it is extremely difficult to 
find {\em designing} sequences
(which possess a nondegenerate ground state) with the standard HP model (cf.\ Table~\ref{tab:desseq}), 
in particular for comparatively long sequences. 
For a statistical analysis of the folding behavior of
this 42-mer, the density of states, as has already been shown in Fig.~\ref{fig:42iter}, 
as well as thermodynamic quantities and their fluctuations were calculated~\cite{bj1}. 
In Fig.~\ref{fig:42fl}, the specific heat $C_V$ and fluctuations of the structural quantities
radius of gyration $R_{\rm gyr}$ and end-to-end distance $R_{\rm ee}$
are plotted as functions of temperature. Two temperature regions of conformational activity 
(shaded in gray),
where the curves of the fluctuating quantities exhibit extremal points, 
can clearly be separated. 

\begin{figure}[t]
\centerline {
\epsfxsize=11cm \epsfbox {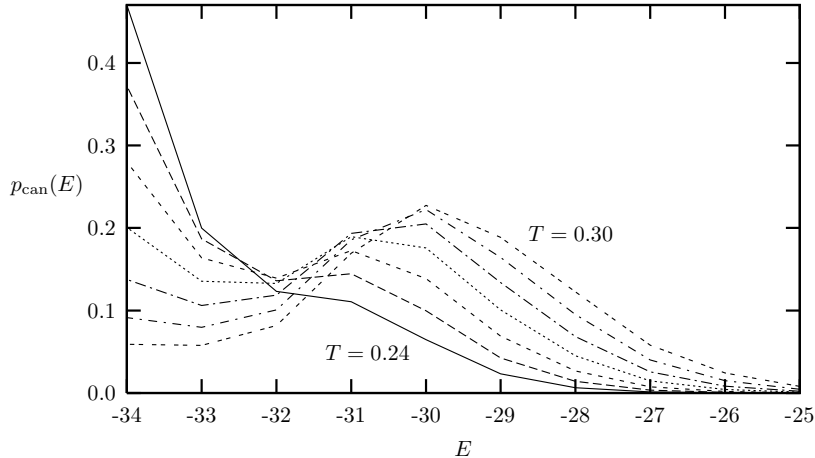}
}
\caption{\label{fig:42pE} 
Canonical energy distributions of the 42-mer for temperatures $T=0.24,0.25,\ldots,0.30$ 
close to the hydrophobic-core collapse transition.
}
\end{figure}

For high temperatures, random conformations are favored. In consequence, in the corresponding, rather 
entropy-dominated ensemble, the high-degenerate high-energy structures govern the thermodynamic 
behavior of the macrostates. A typical representative is shown as an inset in the 
high-temperature pseudophase in Fig.~\ref{fig:42fl}. Annealing the system (or, equivalently,
decreasing the solvent quality), the heteropolymer experiences a conformational transition
towards globular macrostates. A characteristic feature of these intermediary ``molten'' globules
is the compactness of the dominating
conformations as expressed by a small gyration radius. Nonetheless, the conformations
do not exhibit a noticeable internal long-range symmetry and behave rather like a fluid.
Local conformational changes are not hindered by strong free-energy barriers. The
situation changes by entering the low-temperature (or poor-solvent) conformational
phase. In this region, energy dominates over entropy and the effectively attractive 
hydrophobic force favors the formation of a maximally compact core of hydrophobic monomers. 
Polar residues are expelled to the surface of the globule and form a shell that
screens the core from the (fictitious) aqueous environment. 
In Fig.~\ref{fig:42pE}, we have plotted canonical energy distributions
$p_{\rm can}(E)$ for several temperatures near the hydrophobic-core collapse
transition. For temperatures above the transition region (which is between
$T^{(1)}=0.24$ and $T^{(2)}=0.28$, cf.\ Fig.~\ref{fig:42fl}), globular conformations
are more probable, whereas for smaller temperatures hydrophobic-core states dominate.
From the two-peak structure of the distributions in the transition region it can be
concluded that this transition is first-order-like, i.e., both types of macrostates
coexist in this temperature region. 

The existence of the
hydrophobic-core collapse renders the folding behavior of a heteropolymer 
different from crystallization or amorphous transitions of homopolymers. The reason
is the disorder induced by the sequence of different monomer types. The hydrophobic-core
formation is the main cooperative conformational transition which accompanies
the tertiary folding process of a single-domain protein.    

A very important aspect in the discussion of ground-state properties and conformational
transitions towards the native fold is the influence of the heteropolymer sequence.  
For this purpose, we analyze ten designed sequences with 48 monomers, listed in
Table~\ref{tab:48mers}, as given in Ref.~\cite{dill3}. The ratio between the 
numbers of hydrophobic and 
polar residues is one half for these HP proteins, i.e., the  hydrophobicity 
is $n_H=24$. The minimum energies we found from multicanonical chain-growth
simulations~\cite{bj2} coincide with the
values given in Refs.~\cite{dill3,hsu1}. 
Also listed in Table~\ref{tab:48mers} are the estimates for the degeneracies $g_0$ 
of the respective ground-state energies. For comparison, previously given
lower bounds $g^<_{\rm CHCC}$~\cite{shak2} are listed, which were
obtained by means of the constraint-based hydrophobic core construction (CHCC) 
method~\cite{dill3}. Utilizing the idea of a highly compact hydrophobic core
in the native fold, the hydrophobic monomers are in this method arranged in
frames of maximal compactness. The number of the associated so-called Hamiltonian walks that connect the
monomers, respecting the nonchangeable HP sequence, gives a lower bound for the
degeneracy of the ground state. If, due to the sequence, a matching walk cannot be
constructed, the compactness of the frame is relaxed and the search starts anew.
The exact ground-state degeneracy would be obtained by scanning all frames and
searching for conformations with the ground-state energy. Since the method
is of exact enumeration type, the efforts of determining the precise 
ground-state degeneracy are enormous and, therefore, the main power of
this method lies in the identification of native folds and the possibility 
to give a lower bound for its degeneracy.

\begin{figure}[t]
\centerline {
\epsfxsize=11.5cm \epsfbox {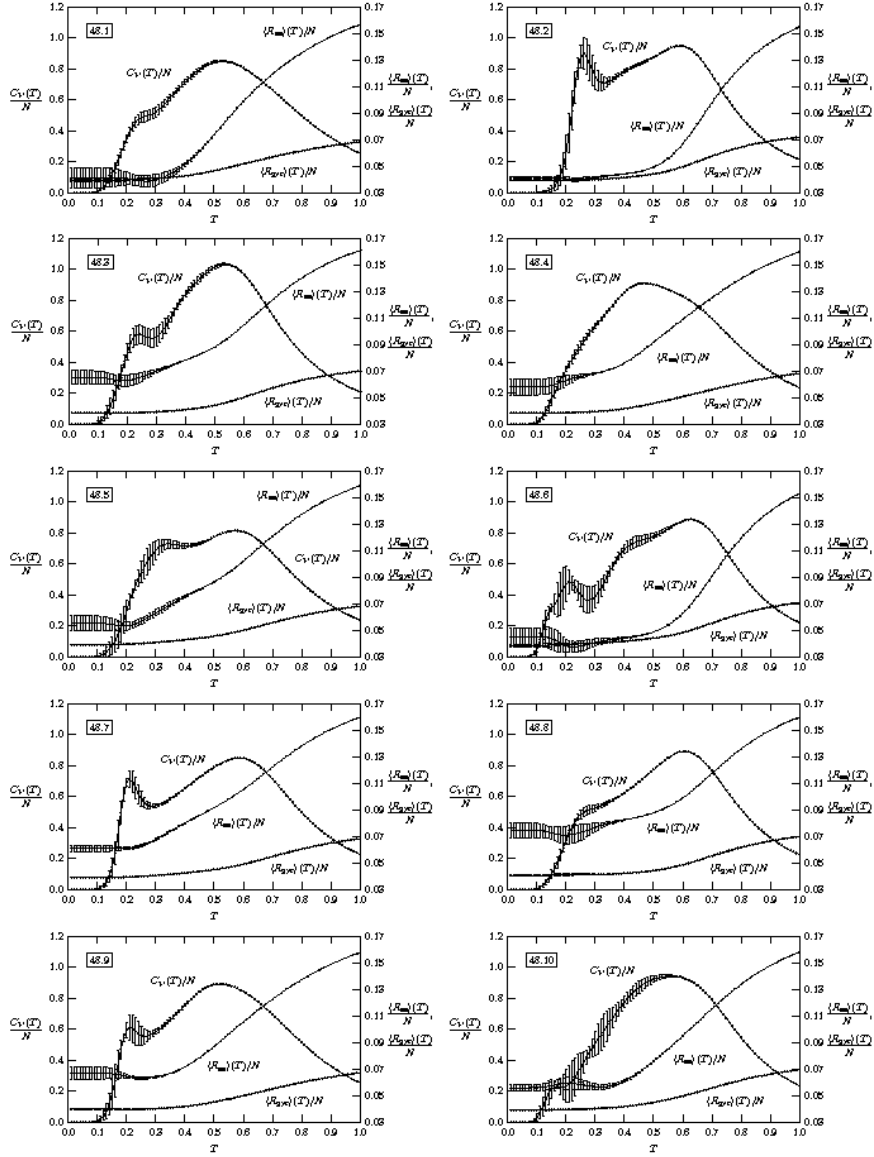}
}
\caption{\label{fig:48all} 
Specific heat, mean radius of gyration, and mean end-to-end distance for the ten
48-mers listed in Table~\ref{tab:48mers}~\cite{bj2}. 
}
\end{figure}

\begin{table}[t]
\caption{\label{tab:48mers} Ground-state energies $E_{\rm min}$ and degeneracies $g_0$
as estimated with the multicanonical chain-growth method~\cite{bj1,bj2} for ten HP sequences with 
48 monomers. For comparison, we have also quoted the lower bounds on native degeneracies
$g^<_{\rm CHCC}$ obtained by means of the CHCC (constraint-based hydrophobic core construction)
method~\cite{dill3} as given in Ref.~\cite{shak2}. In both cases the constant factor $48$ from
rotational and reflection symmetries of conformations spreading into all three spatial directions
was divided out.}
\centerline{\epsfxsize=12cm \epsfbox{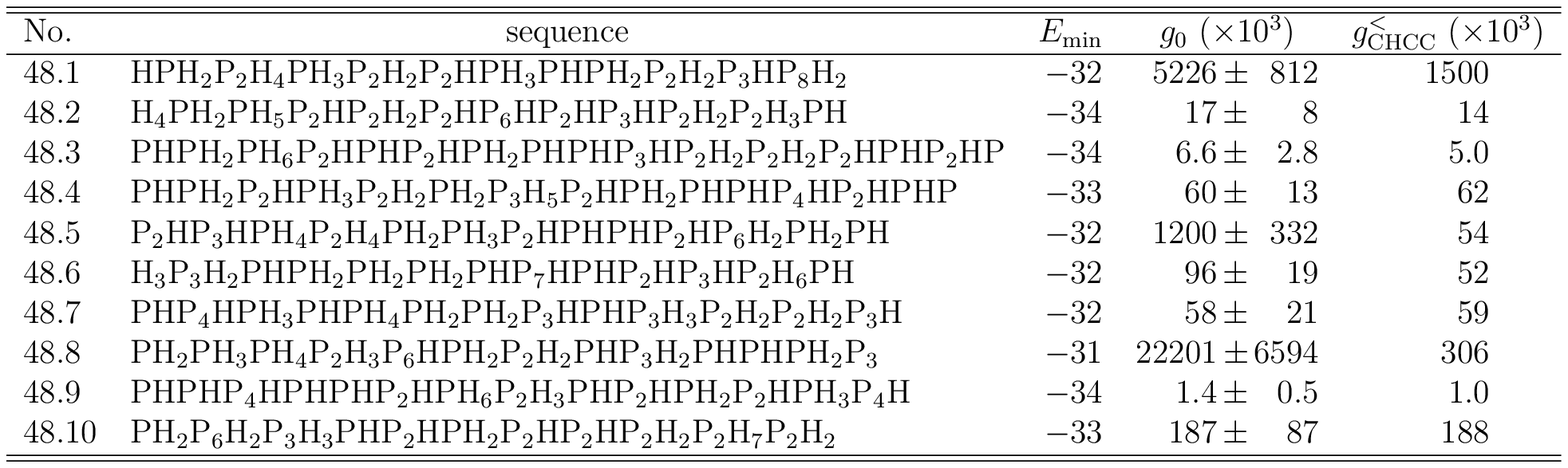}}
\end{table}
For the 48-mers, the $g_0$ values obtained within the multicanonical chain-growth simulation 
lie indeed above these lower bounds or include it within the range of statistical
errors. Notice that for the sequences 48.1, 48.5, and 48.8, the estimates for the ground-state 
degeneracy are much higher than the bounds $g^<_{\rm CHCC}$. In these cases the
smallest frame containing the entire hydrophobic core is rather large 
(cube containing $4\times 3\times 3=36$ monomers with surface area 
$A=32$ [bond length]$^2$) such that 
enumeration of this frame is cumbersome. For 48.5 and 48.8, we further found 
ground-state conformations lying in less compact frames 
(48.5: $A=32,40,42,48,52,54$ [bond length]$^2$, 
48.8: $A=32,40,42$ [bond length]$^2$) and those conformations would require still more
effort to be identified with the CHCC algorithm, which was designed to locate 
global energy minima and therefore starts the search with the most compact 
hydrophobic frames. The ground-state energies of these examples are rather high 
($E_{\rm min}=-31$ for 48.8, and $E_{\rm min}=-32$ for 48.1 and 48.5) and therefore a
higher degeneracy seems to be natural. This is, however, only true, if there does not
exist a conformational barrier that separates the compact H-core low-energy states
from the general compact globules. Comparing the ground-state degeneracies
and the low-temperature behavior of the specific heats for the sequences 48.1, 48.5, 48.6, and
48.7 (all of them having global energy minima with $E_{\rm min}=-32$) as shown 
in Fig.~\ref{fig:48all}, we observe
that 48.6 and 48.7 with rather low ground-state degeneracy actually possess a 
pronounced low-temperature 
peak in the specific heat, while the higher-degenerate proteins 48.1 and 48.5 only 
show up a weak indication of a structural transition at low temperatures. The HP proteins
48.2, 48.3, and 48.9, which have the lowest minimum energy $E_{\rm min}=-34$ among the examples
in Table~\ref{tab:48mers}, have also the lowest ground-state degeneracies. 
These three candidates seem indeed to exhibit a rather strong ground-state -- globule transition, 
as can be read off from the associated specific heats in Fig.~\ref{fig:48all}. 

In Fig.~\ref{fig:48all}, also the mean end-to-end distances and mean radii of gyration are  
plotted as functions of temperature. Both quantities 
usually serve to interpret the conformational compactness of polymers. For HP proteins,
the end-to-end distance is strongly influenced, however, by the types of monomers
attached to the ends of the chain. It is easily seen from the figures that the
48mers with sequences starting and ending with a hydrophobic residue (48.1, 48.2, and 48.6)
have a smaller mean end-to-end distance at low temperatures than the other examples from 
Table~\ref{tab:48mers}. The reason is that the ends can form hydrophobic contacts and therefore
a reduction of the energy can be achieved. Thus, in these cases contacts between ends are 
usually favorable and the mean end-to-end distance is close to the mean radius
of gyration. Interestingly, there exists indeed a crossover region, where 
$\langle R_{\rm ee}\rangle < \langle R_{\rm gyr}\rangle$. Comparing with the behavior 
of the specific heat, this interval is close to the region, where the 
phase dominated by low-energy states crosses over to the globule-favored phase. 
The hydrophobic contact between the ends is strong enough to resist the thermal fluctuations
in that temperature interval. The reason is that, once such a hydrophobic contact between the ends is
established, usually other in-chain hydrophobic monomers are attracted and form a
hydrophobic core surrounding the end-to-end contact. Thus, before the contact
between the ends is broken, an increase of the temperature first leads to a melting
of the surrounding contacts. 
The entropic freedom to form new conformations 
is large since the low-energy states are all relatively high degenerate
and do not possess symmetries requiring an appropriate amount of heat to be broken.
For sequences possessing mixed or purely polar ends, the mean end-to-end distance and 
mean radius of gyration differ much stronger, as there is no energetic reason, why 
the ends occupy nearest-neighbor positions. 

In conclusion, we see that for longer chains the strength of the low-temperature transition
not only depends on low ground-state degeneracies as it does for short chains~\cite{bj1}. Rather, 
the influence of the higher-excited states cannot be neglected. A striking example is
sequence 48.4 with rather low ground-state degeneracy, but only weak signals for a 
low-temperature transition.
\subsection{Specificity of protein adsorption to selective solid substrates}
\index{transition!adsorption}
\label{ssec:hpad}
\index{adsorption}
In this section, we discuss results of a simple lattice model similar to Eq.~(\ref{eq:adsmodel}) 
for analyzing the conformational behavior of HP proteins in adsorption processes to different, 
specific solid substrates. The objective is the determination of a pseudophase\index{pseudophase}
diagram, which allows for the classification of conformational subphases
in dependence of the external parameters temperature and solubility of the surrounding
(implicit) solvent.

\begin{figure}[t]
\centerline {
\epsfxsize=6cm \epsfbox {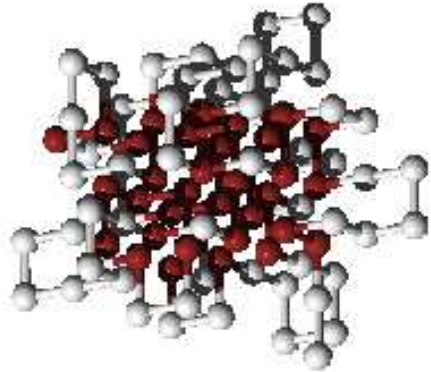}
}
\caption{\label{fig:103mer} 
Compact hydrophobic-core conformation of the 103-mer~\cite{bj2} used in the 
peptide adsorption study~\cite{bj5,bj6}. Dark spheres correspond to hydrophobic
monomers and light spheres mark polar residues.
}
\end{figure}
The recent developments in single molecule experiments at the nanometer
scale, e.g., by means of atomic force microscopy (AFM)~\cite{afm1} and optical
tweezers~\cite{ot1}, allow now for a more detailed exploration of structural properties
of polymers in the vicinity of adsorbing substrates~\cite{gray1}. The possibility to perform
such studies is of essential biological and technological significance. From
the biological point of view the understanding of the binding and docking mechanisms of
proteins at cell membranes is important for the reconstruction of biological
cell processes. Similarly, specificity of peptides and binding affinity
to selected substrates could be of great importance for future electronic nanoscale
circuits and pattern recognition nanosensory devices~\cite{nakata1}. The study of
hybrid interface models has considerable applications for
a broad variety of problems, e.g., understanding the mechanisms of protein--ligand
binding~\cite{ligand1}, prewetting and layering transitions in polymer solutions
as well as dewetting of polymer films~\cite{wetting1,binder1}, molecular pattern
recognition~\cite{bogner1}, electrophoretic polymer deposition and growth~\cite{foo1}.
Recently, the influence of adhesion and steric hindrance for
polymers grafted to a flat substrate~\cite{grass4,vrbova,singh,causo1,prellberg1,huang1},
conformational pseudophase transitions\index{transition!conformational} for nongrafted polymers and peptides in a 
cavity with attractive substrate~\cite{bj3,bj4,bj5,bj6}, the shape response to
pulling forces~\cite{celestini1,prellberg2} or external
fields~\cite{frey1} were subject of computer simulations and analytical approaches
of different models.
The question how
a flexible substrate, e.g., a cell membrane, bends as a reaction of a grafted polymer,
was, for example, addressed in Ref.~\cite{lipowsky1}. Proteins exhibit a strong
specificity as the affinity of peptides to adsorb at surfaces depends on the amino
acid sequence, solvent properties, and substrate shape. 
This was experimentally and numerically studied, e.g.,
for peptide-metal~\cite{brown1,schulten1} 
and peptide-semiconductor~\cite{whaley1,goede1} interfaces.
Binding/folding and docking properties of lattice heteropolymers at an adsorbing
surface were also subject of numerical studies~\cite{irbaeck1}.
\subsubsection{Lattice model for hybrid peptide-substrate interfaces}
\begin{figure}[t]
\centerline {
\epsfxsize=10cm \epsfbox {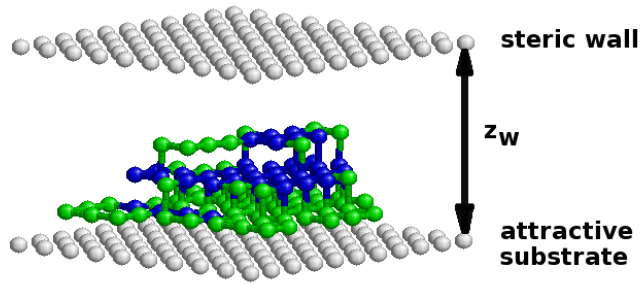}
}
\caption{\label{fig:adsmod} 
Lattice model used in the peptide-substrate adsorption study.
}
\end{figure}
For the study of hybrid peptide-substrate models, we use the HP transcription of the 103-residue 
protein {\em cytochrome c}, which was extensively
studied in the past~\cite{hsu1,bj2,103lat,103toma}. The HP sequence
contains 37 hydrophobic and 66 polar residues. A conformation with a
highly compact hydrophobic-core, exhibiting 56 hydrophobic contacts, 
is shown in Fig.~\ref{fig:103mer}. 
This lattice peptide 
resides in a cavity with an attractive substrate (see Fig.~\ref{fig:adsmod}). For
regularization of the upper halfspace, an additional steric wall in a distance $z_w$
is introduced. The value of $z_w$ is chosen sufficiently large to keep the influence 
on the unbound heteropolymer small (in the actual example, $z_w=200$ was used).  
In order to study the 
specificity of residue binding, we distinguish three
substrates with different affinities to attract the peptide monomers: 
(a) the type-independent attractive, (b) 
the hydrophobic, and (c) the polar substrate. The number of corresponding 
nearest-neighbor contacts
between monomers and substrate shall be denoted as $n_s^{H+P}$, $n_s^{H}$, 
and $n_s^{P}$, respectively.
In analogy to the polymer-substrate model~(\ref{eq:adsmodel}), 
we express the energy of the hybrid peptide-substrate system simply by 
\begin{equation}
\label{eq:energy}
E_s(n_s,n_{\rm HH})=-\varepsilon_0(n_s+sn_{\rm HH}),
\end{equation} 
where $n_s=n_s^{H+P}$, $n_s^{P}$, or $n_s^{H}$ depending on the substrate 
(we set $\varepsilon_0=1$ in the following). 
The solubility (or reciprocal solvent 
parameter) $s$ is, as well as the temperature $T$, an external parameter. It controls 
the quality of the solvent
(the larger the value of $s$, the worse the solvent). This model was investigated
by means of the contact-density chain-growth algorithm\index{chain-growth!contact density} (see Sec.~\ref{ssec:cgmeth}), 
which allows a direct estimation 
of the degeneracy (or density) $g(n_s, n_{\rm HH})$
of macrostates of the system with given contact numbers $n_s$ and 
$n_{\rm HH}$~\cite{bj5,bj6}.  
In contrast to move-set based Metropolis Monte Carlo or conventional chain-growth
methods which would require many separate simulations to obtain
results for different parameter pairs $(T,s)$ and which frequently suffer from slowing down
in the low-temperature sector, the contact-density chain-growth method allows the 
computation of the {\em complete}
contact density for each system within a {\em single} simulation run. Since the contact
density is independent of temperature and solubility, energetic quantities
such as the specific heat~(\ref{eq:adssh}) can easily be calculated for all 
values of $T$ and $s$. Nonenergetic
quantities require accumulated densities to be measured within the simulation, but this is also
no problem.
\subsubsection{Conformational adsorption behavior in dependence of temperature
and solubility}
\begin{figure}
\centerline {
\epsfxsize=12cm \epsfbox {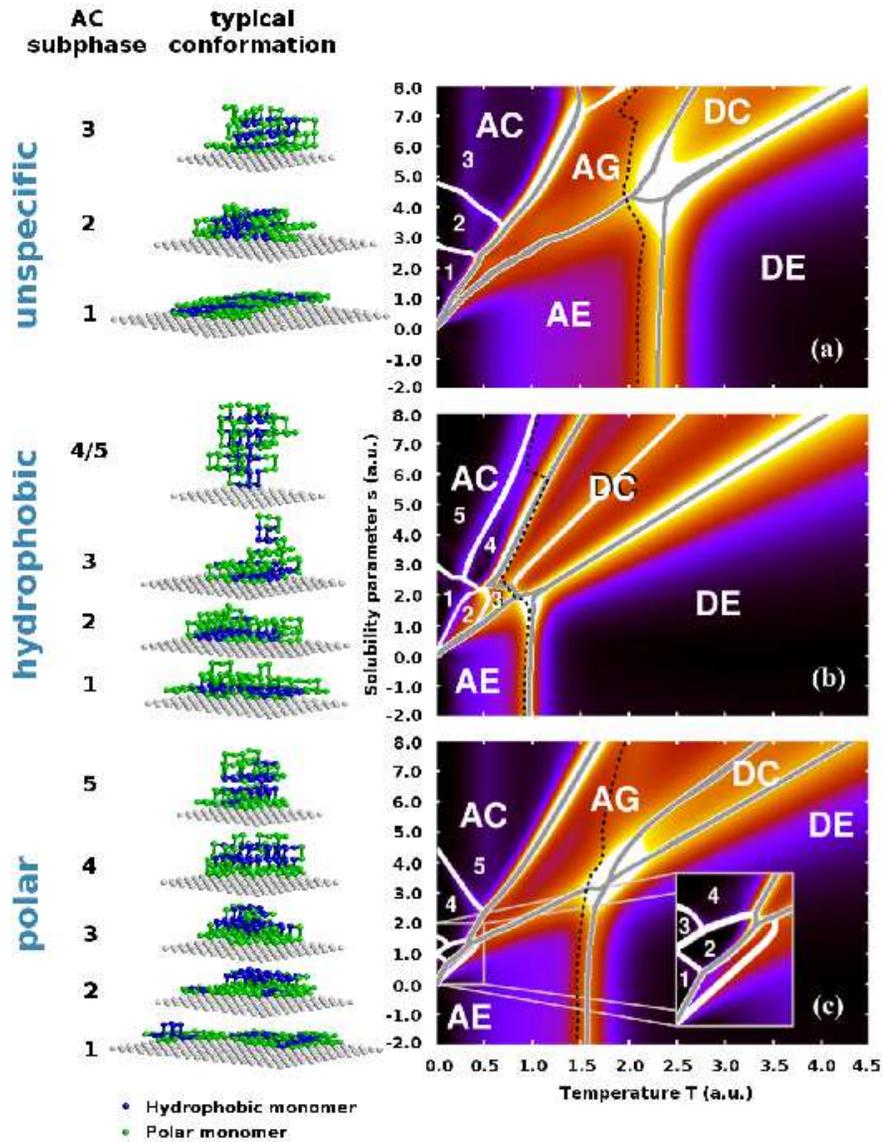}
}
\caption{\label{fig:pd103} 
Specific-heat profiles as a function of temperature $T$ and solubility
parameter $s$ of the 103-mer near three different substrates that are attractive for
(a) all, (b) only hydrophobic, and (c) only polar monomers.
White lines indicate the ridges of the profile. Gray lines mark the
main ``phase boundaries''. The dashed black line represents the first-order-like
binding/unbinding transition state, where the contact free energy possesses two minima
(the adsorbed and the desorbed state). In the left panel typical conformations
dominating the associated AC phases of the different systems are shown.
}
\end{figure}
In Figs.~\ref{fig:pd103}(a)--(c) the color-coded profiles of the specific heats for the different
substrates are shown (the brighter the larger the value of $C_V$). We interpret the ridges (for accentuation marked
by white and gray lines) as the boundaries of the pseudophases. The gray lines indicate the
main transition lines, while the white lines separate pseudophases that strongly depend on specific properties 
of the heteropolymer, such as its exact number and sequence of hydrophobic and polar monomers. 
With its degeneracy  $g(n_s,n_{\rm HH})$, we define the contact free energy as 
$F_{T,s}(n_s,n_{\rm HH}) = E_s(n_s,n_{\rm HH}) -T \ln\, g(n_s,n_{\rm HH})$ and 
the probability for a macrostate with $n_s$ substrate and $n_{\rm HH}$ hydrophobic contacts as
$p_{T,s}(n_s,n_{\rm HH})\sim g(n_s,n_{\rm HH})\exp(-E_s/T)$. 
Assuming that the minimum of the free-energy landscape $F_{T,s}(n_s^{(0)},n_{\rm HH}^{(0)})\to {\rm min}$ for 
given external parameters $s$ and $T$ is related to the class of macrostates with $n_s^{(0)}$ surface and 
$n_{\rm HH}^{(0)}$ hydrophobic contacts, this class dominates the phase the system resides in. For this
reason, it is instructive to calculate all minima of the contact free energy and to determine the 
associated contact numbers in a wide range of values for the external parameters. 

\begin{figure}
\centerline{\epsfxsize=12cm \epsfbox{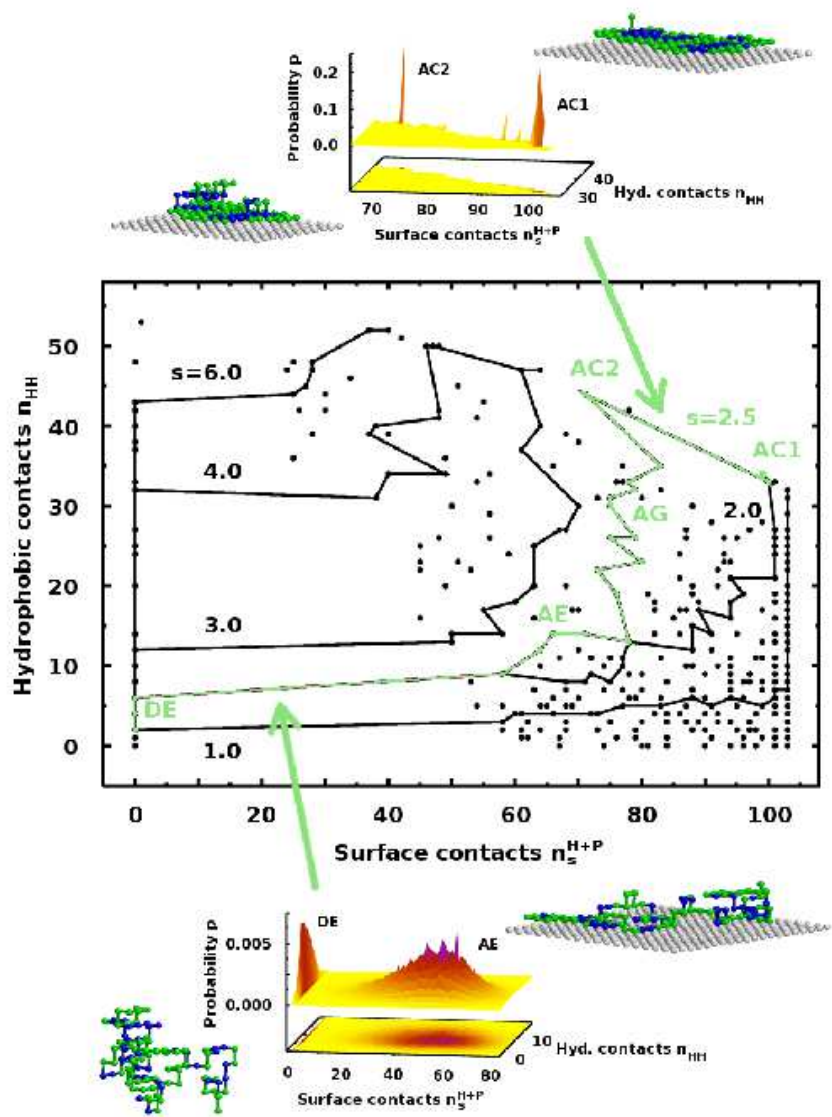}}
\caption{\label{fig:lm} Contact-number map of all free-energy minima for the 103-mer
and substrate equally attractive to all monomers. Full circles correspond to minima of the contact free energy 
$F_{T,s}(n_s^{\rm H+P},n_{\rm HH})$
in the parameter space $T\in [0,10]$, $s\in[-2,10]$. Lines illustrate how the contact free energy changes
with the temperature at constant solvent parameter $s$. For the exemplified
solvent with $s=2.5$, the peptide experiences near $T=0.35$ a sharp first-order-like layering transition
between single- to double-layer conformations (AC1,2). Passing the regimes of adsorbed globules (AG)
and expanded conformations (AE), the discontinuous binding/unbinding transition from AE to DE happens
near $T=2.14$. In the DE phase the ensemble is dominated by desorbed, expanded conformations.
Representative conformations of the phases are shown next to the respective peaks of the probability
distributions.
}
\end{figure}

The map of all possible free-energy 
minima in the range of external parameters $T\in[0,10]$ and $s\in[-2,10]$ is shown in Fig.~\ref{fig:lm} for
the peptide in the vicinity of a substrate that is equally attractive for both hydrophobic and polar monomers. 
Solid lines visualize ``paths'' through the free energy landscape when changing temperature under constant solvent 
($s={\rm const}$) conditions. Let us follow the exemplified trajectory for $s=2.5$. 
Starting at very low temperatures, 
we know from the pseudophase diagram in Fig.~\ref{fig:pd103}(a) that the system resides in pseudophase AC1. This means
that the macrostate of the peptide is dominated by the class of compact, film-like single-layer conformations. The system
obviously prefers surface contacts at the expense of hydrophobic contacts. Nonetheless, the formation of compact
hydrophobic domains in the two-dimensional topology is energetically favored but maximal compactness is hindered by the steric
influence of the substrate-binding polar residues. Increasing the temperature, the system experiences close to $T\approx 0.35$
a sharp first-order-like conformational transition, and a second layer forms (AC2). This is a mainly entropy-driven transition
as the extension into the third dimension perpendicular to the substrate surface increases the number of possible 
peptide conformations. Furthermore, the loss of energetically favored substrate contacts of polar monomers is 
partly compensated by the energetic gain due to the more compact hydrophobic domains. Increasing the temperature
further, the density of the hydrophobic domains reduces and overall compact conformations dominate in the globular
pseudophase AG. Reaching AE, the number of hydrophobic contacts decreases further, and also the total number of
substrate-contacts. Extended, dissolved conformations dominate. The transitions from AC2 to AE via AG are comparatively 
``smooth'', i.e., no immediate
changes in the contact numbers passing the transition lines are noticed. Therefore, these conformational transitions
could be classified as second-order-like. The situation is different when approaching the unbinding
transition line from AE  close to $T\approx 2.14$. This transition is accompanied by a dramatic loss of substrate 
contacts -- the peptide desorbs from the substrate and behaves in pseudophase DE like a free peptide, i.e., the substrate
and the opposite neutral wall regularize the translational degree of freedom perpendicular to the walls, but rotational
symmetries are unbroken (at least for conformations not touching one of the walls). As the probability distribution
in Fig.~\ref{fig:lm} shows, the unbinding transition is also first-order-like, i.e., close to the transition line, there
is a coexistence of adsorbing and desorbing classes of conformations. 

Despite the surprisingly rich and complex phase behavior there are main ``phases'' that can be
distinguished in all three systems. 
These are separated in Figs.~\ref{fig:pd103}(a)--(c) by gray lines.
Comparing the three systems we find that they all possess
pseudophases, where adsorbed compact (AC), adsorbed expanded (AE), desorbed compact (DC), and
desorbed expanded (DE) conformations dominate. ``Compact'' here 
means that the heteropolymer has formed a 
dense hydrophobic core, while expanded conformations are dissolved, random-coil like.
The sequence and substrate specificity of
heteropolymers generates, of course, a rich set of new interesting and selective phenomena
not available for homopolymers. One example is the pseudophase of adsorbed globules (AG), which is noticeably present
only in those systems, where all monomers are equally attractive to the substrate (Fig.~\ref{fig:pd103}(a)) and
where polar monomers favor contact with the surface (Fig.~\ref{fig:pd103}(b)). In this phase, the conformations
are intermediates in the binding/unbinding region. This means that monomers 
currently desorbed from the substrate have not yet found their position within a compact conformation.
Therefore, the hydrophobic core, which is smaller than in the respective adsorbed phase 
(i.e., at constant solubility $s$), appears as a loose cluster of hydrophobic monomers. 
\begin{figure}
\centerline{\epsfxsize=12cm \epsfbox{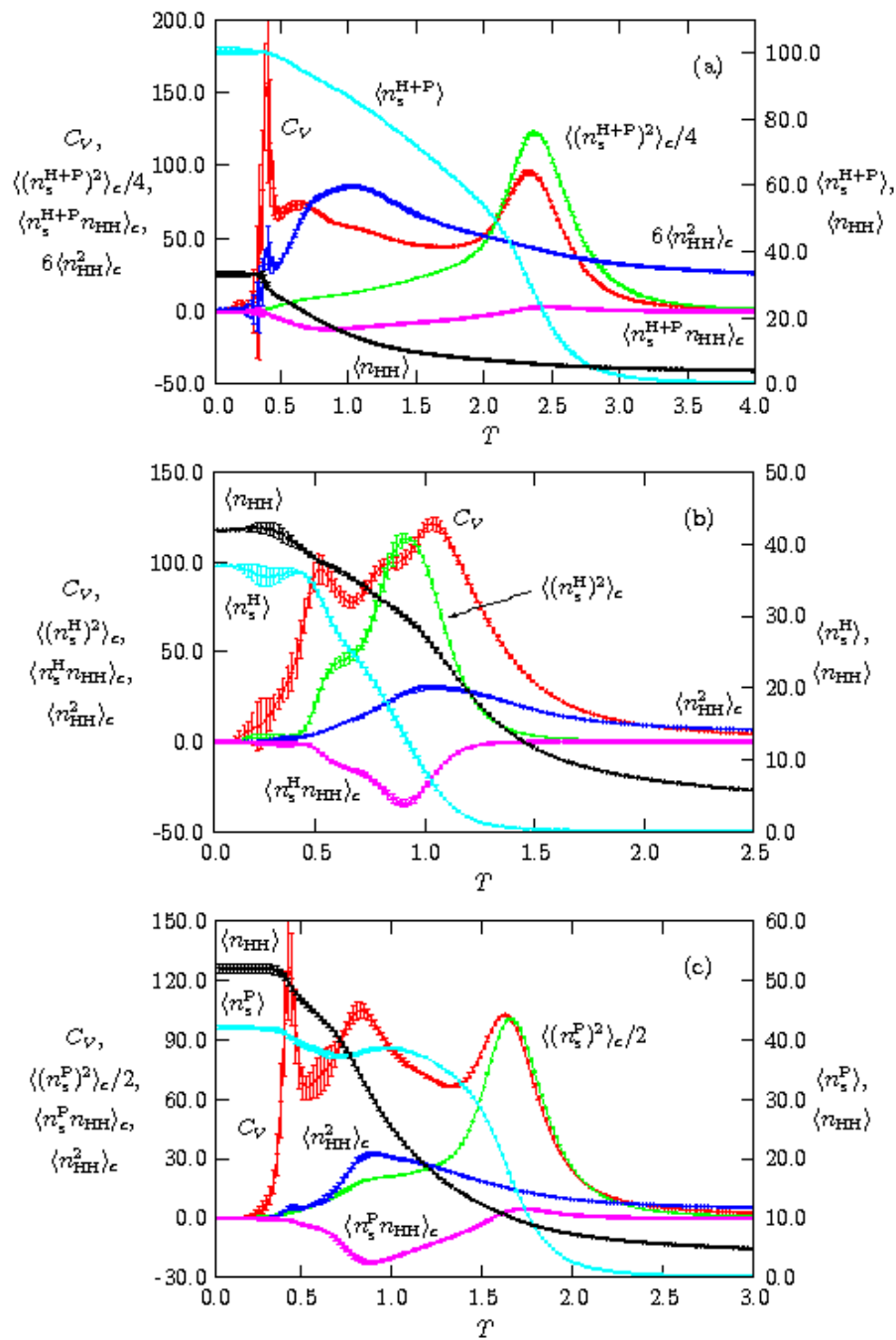}}
\caption{\label{fig:c103}  
Temperature dependence of specific heat, correlation matrix components, and 
contact number expectation values of the 103mer for surfaces attractive for 
(a) all, (b) only hydrophobic, and (c) only polar monomers at $s=2$. 
}
\end{figure}

In Figs.~\ref{fig:c103}(a)--(c), we have plotted, exemplified for $s=2$,  
the statistical averages of the contact numbers $n_s$ and $n_{\rm HH}$
as well as their variances and covariances for the three systems. For comparison 
we have also included the specific heat, 
whose peaks correspond to the intersected transition lines of
Figs.~\ref{fig:pd103}(a)--(c) at $s=2$. From Figs.~\ref{fig:c103}(a) and (c) we read off that the
transition from AC to AG near $T\approx 0.4$ is mediated by fluctuations of the 
intrinsic hydrophobic contacts. The very dense hydrophobic domains in the AC subphases
lose their compactness. This transition is absent in the hydrophobic-substrate system
(Fig.~\ref{fig:c103}(b)). The signal seen belongs to a hydrophobic layering AC subphase transition,
which influences mainly the number of surface contacts $n_s^{\rm H}$.
The second peak of the specific heats belongs to the transition between adsorbed compact or
globular (AC/AG) and expanded (AE) conformations. This behavior is similar in all
three systems. Remarkably, it is accompanied by a strong anti-correlation between 
surface and intrinsic contact numbers, $n_s$ and $n_{\rm HH}$. Not surprisingly,
the hydrophobic contact number $n_{\rm HH}$ fluctuates stronger than the number
of surface contacts, but apparently in a different way. Dense conformations with
hydrophobic core (and therefore many hydrophobic contacts) possess a relatively small
number of surface contacts. Vice versa, conformations with many surface contacts
cannot form compact hydrophobic domains. Finally, the third specific heat peak 
marks the binding/unbinding transition, which is, as expected, due to a strong
fluctuation of the surface contact number.

The strongest difference between the three systems is their behavior in pseudophase AC, which is
roughly parameterized by $s>5T$.
If hydrophobic and polar monomers are equally attracted by the substrate (Fig.~\ref{fig:pd103}(a)),
we find three AC subphases in the parameter space plotted. In subphase AC1, film-like conformations
dominate, i.e., all 103 monomers are in contact with the substrate. Due to the good solvent quality in 
this region, the formation of a hydrophobic core is less attractive than the maximal deposition of all monomers 
at the surface, the ground state is $(n_s^{\rm H+P},n_{\rm HH})_{\rm min}=(103,32)$. In fact, instead of
a single compact hydrophobic core there are nonconnected hydrophobic clusters. At least on the used
simple cubic lattice and the chosen sequence, the formation of a single hydrophobic core is necessarily
accompanied by an unbinding of certain polar monomers and, in consequence, an extension of the
conformation into the third spatial dimension. In fact, this happens when entering AC2 
[$(n_s^{\rm H+P},n_{\rm HH})_{\rm min}=(64,47)$], 
where a single hydrophobic
two-layer domain has formed at the expense of losing surface contacts. In AC3,
the heteropolymer has maximized the number of hydrophobic contacts and only local arrangements of monomers 
on the surface of the very compact structure lead to the still possible maximum number of substrate
contacts. $F_{T,s}$ is minimal for $(n_s^{\rm H+P},n_{\rm HH})_{\rm min}=(40,52)$.

The behavior of the heteropolymer adsorbed at a surface that is only attractive to
hydrophobic monomers (Fig.~\ref{fig:pd103}(b)) is apparently different in the AC phase. 
Since surface contacts 
of polar monomers are energetically not favored, the subphase structure is determined
by the competition of two hydrophobic forces: substrate attraction and formation of intrinsic
contacts. In AC1, the number of 
hydrophobic substrate contacts is maximal for the single 
hydrophobic layer, $(n_s^{\rm HH},n_{\rm HH})_{\rm min}=(37,42)$. The {\em single} two-dimensional hydrophobic domain is also
maximally compact, at the expense of displacing polar monomers into a second layer.
In subphase AC2 intrinsic contacts are entropically 
broken with minimal free energy for $35\le n_{\rm HH}\le 40$, while $n_s^{\rm HH}=37$
remains maximal. 
Another AC subphase, AC3, exhibits a
hydrophobic layering transition at the expense of hydrophobic substrate contacts. Much more interesting
is the subphase transition from AC1 to AC5. The number of hydrophobic substrate 
contacts $n_s^{\rm HH}$ of the ground-state conformation dramatically decreases 
(from $37$ to $4$) and the hydrophobic monomers collapse
in a one-step process from the compact two-dimensional domain to the maximally compact
three-dimensional hydrophobic core. The conformations are mushroom-like structures grafted 
at the substrate. AC4 is similar to AC5, with advancing desorption.

Not less exciting is the subphase structure of the heteropolymer interacting with a polar substrate
(Fig.~\ref{fig:pd103}(c)).
For small values of $s$ and $T$, the behavior of the heteropolymer is dominated by the competition
between polar monomers contacting the substrate and hydrophobic monomers favoring the formation
of a hydrophobic core, which, however, also requires cooperativity of the polar monomers.
In AC1, film-like conformations ($n_s^{\rm P}=66$, $n_{\rm HH}=31$)
with disconnected hydrophobic clusters dominate. Entering AC2, hydrophobic contacts 
are energetically favored and a second hydrophobic layer forms at the expense
of a reduction of polar substrate contacts 
[$(n_s^{\rm P},n_{\rm HH})_{\rm min}=(61,37)$]. 
In AC3, the upper layer 
is mainly hydrophobic [$(n_s^{\rm P},n_{\rm HH})_{\rm min}=(53,45)$], 
while the poor quality of the solvent ($s$ large) and the comparatively 
strong hydrophobic force let the conformation further collapse [AC4: $(n_s^{\rm P},n_{\rm HH})_{\rm min}=(42,52)$]
and the steric cooperativity forces more 
polar monomers to break the contact to the surface and to form a shell surrounding the hydrophobic
core [$(n_s^{\rm P},n_{\rm HH})_{\rm min}=(33,54)$ in AC5]. 
\section{Going off-lattice: Folding behavior of heteropolymers in the AB continuum model}
\label{sect:ab}
The lattice models discussed in the previous sections suffer from the fact that the results 
for the finite-length heteropolymers typically depend on the underlying lattice type. It is difficult
to separate realistic effects from artefacts induced by the use of a certain lattice structure. This
problem can be avoided, in principle, by studying off-lattice heteropolymers, where the 
degrees of freedom are continuous. On the other hand, this advantage is partly counter-balanced by the 
increasing computational efforts for sampling the relevant regions of the conformational state space.
In consequence, a precise analysis of statistical properties of off-lattice heteropolymers by means
of generalized-ensemble methods can reliably be performed only for chains much shorter than those
considered in the lattice studies. In the following, we focus on hydrophobic-polar heteropolymers
with 20 monomers employing the so-called AB model~\cite{still1}\index{AB model}, where $A$ monomers are hydrophobic and residues of
type $B$ are polar (or hydrophilic). 
\subsection{Modeling and updating}
\label{ssect:model}
\begin{figure}[t]
\centerline{\epsfxsize=8cm \epsfbox{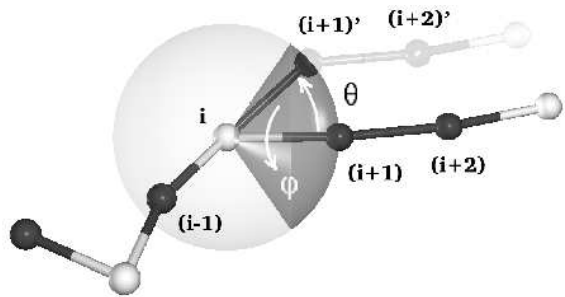}}
\caption{\label{fig:update} Spherical update of the bond vector between the $i$th and
$(i+1)$th monomer.  } 
\end{figure}
We denote the spatial position of the $i$th monomer in a heteropolymer consisting 
of $N$ residues by ${\bf r}_i$, $i=1,\ldots,N$, and the vector connecting nonadjacent monomers $i$ and $j$ 
by ${\bf r}_{ij}$. For covalent bond vectors, we set $|{\bf b}_i|\equiv |{\bf r}_{i\,i+1}|=1$. The bending angle 
between monomers $k$, $k+1$, and $k+2$ is $\vartheta_k$ 
($0\le \vartheta_k\le \pi$) and $\sigma_i=A,B$ symbolizes the type of the monomer. In the
AB model~\cite{still1}, the energy of a conformation is given by
\begin{equation}
\label{eq:ab}
E = \frac{1}{4}\sum\limits_{k=1}^{N-2}(1-\cos \vartheta_k)+
4\sum\limits_{i=1}^{N-2}\sum\limits_{j=i+2}^N\left(\frac{1}{r_{ij}^{12}}
-\frac{C(\sigma_i,\sigma_j)}{r_{ij}^6} \right),
\end{equation} 
where the first term is the bending energy and the sum runs over the $(N-2)$ bending angles of successive 
bond vectors. 
The second term partially competes with the bending barrier by a 
potential of Lennard-Jones type. It depends on the distance between monomers being non-adjacent 
along the chain and accounts for the influence of the AB sequence on the energy.
The long-range behavior is attractive for pairs of like monomers and repulsive for $AB$ pairs
of monomers:
\begin{equation}
\label{stillC}
C(\sigma_i,\sigma_j)=\left\{\begin{array}{cl}
+1, & \hspace{7mm} \sigma_i,\sigma_j=A,\\
+1/2, & \hspace{7mm} \sigma_i,\sigma_j=B,\\
-1/2,  & \hspace{7mm} \sigma_i\neq \sigma_j.\\
\end{array} \right.    
\end{equation}
The Monte Carlo simulation of this model is not straightforward as strictly local
updates are not possible. A simple nonlocal update of a given conformation
can be performed by using the procedure displayed in Fig.~\ref{fig:update}. 
Since the length of the bonds is fixed ($|{\bf b}_k|=1$, $k=1,\ldots,N-1$), the $(i+1)$th monomer
lies on the surface of a sphere with radius unity around the $i$th monomer.
Therefore, spherical coordinates are the natural choice for calculating the new
position of the $(i+1)$th monomer on this sphere. For the reason of efficiency,
we do not select any point on the sphere but restrict the choice to a
spherical cap with maximum opening angle $2\theta_{\max}$ (the dark area in
Fig.~\ref{fig:update}). Thus, to change the position of the $(i+1)$th monomer to
$(i+1)'$, we select the angles $\theta$ and $\varphi$ randomly from the 
respective intervals $\cos \theta_{\max} \le \cos \theta \le 1$ and $0\le\varphi \le 2\pi$,
which ensure a uniform distribution of the $(i+1)$th monomer positions on the associated 
spherical cap.
After updating the position of the
$(i+1)$th monomer, the following monomers in the chain are simply translated 
according to the corresponding bond vectors which remain unchanged in this type
of update. Only the bond vector between the $i$th and the $(i+1)$th monomers is
rotated, all others keep their direction. This is similar to single spin updates
in local-update Monte Carlo simulations of the classical Heisenberg model with the 
difference that in addition to local energy changes long-range interactions of the monomers, changing their relative position 
to each other, have to be computed anew after the update.
For simulations in the state space of dense conformations it is recommendable to choose a rather small opening angle, 
e.g., $\cos\theta_{\max}=0.99$, in order to be able to sample also very narrow and deep valleys in the landscape
of angles.

For the following discussion of folding channels of 20mers~\cite{ssbj1}, these updates were 
used in combination with multicanonical sampling~\cite{muca1,muca2}.
\subsection{Characteristic protein folding\index{protein folding} channels and free-energy landscapes 
from coarse-grained modeling}
\label{ssect:fold}
\begin{table}[t]
\caption{\label{tab:seqs} The three AB 20-mers studied
and the values of the associated (putative) global energy minima. Note that the given values for sequence S3
belong to two different, almost degenerate folds  (cf.\ Fig.~\ref{fig:gemB}).}
\centerline{
\begin{tabular}{cp{3mm}cc}\hline\hline
label & & sequence & global energy minimum~\cite{ssbj1}\\ \hline
S1 & & $BA_6BA_4BA_2BA_2B_2$ & $-33.8236$\\
S2 & & $A_4BA_2BABA_2B_2A_3BA_2$ & $-34.4892$\\
S3 & & $A_4B_2A_4BA_2BA_3B_2A$ & $-33.5838$, $-33.5116$ \\ \hline \hline
\end{tabular}
}
\end{table}
The folding process of proteins is necessarily accompanied by cooperative conformational changes.
Although not phase transitions in the strict sense, it should be expected that one or a few
parameters can be defined that enable the description of the structural ordering process~\cite{du1,pande2}. The 
number of degrees of freedom in most all-atom models is given by the dihedral torsional
backbone and side-chain angles. In coarse-grained C$^\alpha$ models as the AB model used in this study, 
the original dihedral angles are replaced by a set of virtual torsional and bond angles. In fact, the number
of degrees of freedom is not necessarily reduced in simplified off-lattice models. Therefore, 
the complexity of the space of degrees of freedom is comparable with more realistic models, and
it is also a challenge to identify a suitable order parameter for the folding in such minimalistic 
heteropolymer models. 

In analogy to studies of the specific folding behavior in all-atom protein models~\cite{okamoto1,okamoto2}, 
it is suitable to define a generalized variant of an angular overlap order parameter as introduced in Ref.~\cite{baj2}.
The idea is to define a simple and computationally low-cost measure for the similarity of two conformations,
where the differences of the angular degrees of freedom are calculated. In order to consider this 
parameter as kind of order parameter, it is useful to compare conformations ${\bf X}=({\bf r}_1,\ldots,{\bf r}_N)$ 
of the actual ensemble with
a suitable reference conformation, which is preferably chosen to be the global-energy minimum 
conformation ${\bf X}^{(0)}$. We define the overlap parameter as follows:
\begin{equation}
\label{eq:ov}
Q({\bf X})=1 - d({\bf X}). 
\end{equation}
With $N_b=N-2$ and $N_t=N-3$ being the respective numbers of 
bond angles $\Theta_i$ and torsional angles $\Phi_i$,
the angular deviation between the conformations is calculated according to
\begin{equation}
\label{eq:dparam}
d({\bf X})=\frac{1}{\pi(N_b+N_t)}\Bigg[
\sum\limits_{i=1}^{N_b}d_b\left(\Theta_i\right)+
\min_{r=\pm}\left(\sum\limits_{i=1}^{N_t}d_t^r\left(\Phi_i\right)\right)\Bigg],
\end{equation}
where
\begin{eqnarray}
\label{eq:dparamB}
d_b(\Theta_i)&=&|\Theta_i-\Theta^{(0)}_i|,\\
d_t^\pm(\Phi_i)&=&{\rm min} \left(|\Phi_i\pm\Phi^{(0)}_i|,2\pi-|\Phi_i\pm\Phi^{(0)}_i| \right).
\end{eqnarray}
Here it is taken into account that the AB model is invariant under the reflection symmetry
$\Phi_i\to-\Phi_i$. Thus, it is not useful to distinguish between reflection-symmetric
conformations and therefore only the larger overlap is considered.
Since $-\pi\le \Phi_i\le \pi$ and $0\le\Theta_i\le\pi$, the overlap is unity, if all angles 
of the conformations ${\bf X}$ and ${\bf X}^{(0)}$ coincide, else $0\le Q<1$. It should be noted that the average 
overlap of a random conformation with the corresponding reference state is for the sequences considered close to 
$\langle Q\rangle\approx 0.66$.
As a rule of thumb, it can be concluded that values $Q<0.8$ indicate weak or no significant similarity
of a given structure with the reference conformation. 

\begin{figure}[t]
\centerline{\epsfxsize=11.8cm \epsfbox{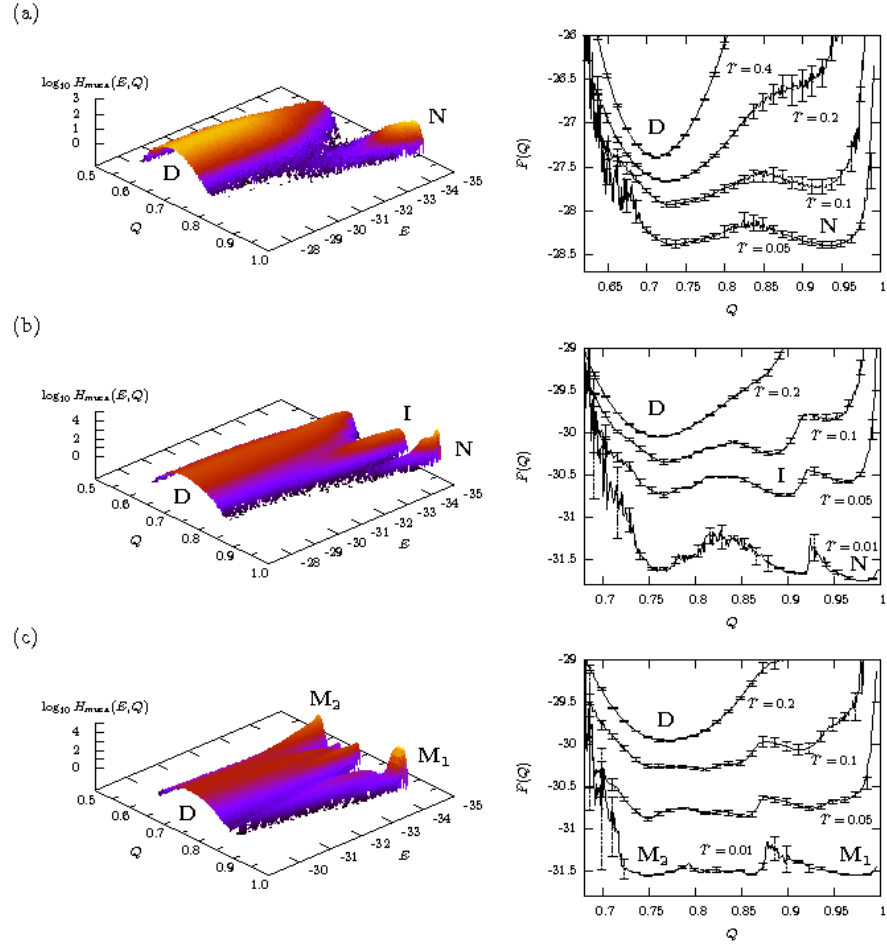}}
\caption{\label{fig:fandh} Multicanonical histograms $H_{\rm muca}(E,Q)$ of energy $E$ and angular overlap parameter $Q$ 
and free-energy landscapes $F(Q)$ at different temperatures for the three sequences (a) S1, (b) S2, and (c) S3. 
The reference folds reside at $Q=1$ and $E=E_{\rm min}$~\cite{ssbj1}.}
\end{figure}
For the qualitative discussion of the folding characteristics, we 
consider the multicanonical histograms of energy $E$ and angular overlap $Q$,
$H_{\rm muca}(E,Q) = \sum_t\,\delta_{E,E({\bf X}_t)}\delta_{Q,Q({\bf X}_t)}$,  
where the sum runs over all Monte Carlo sweeps $t$ in the multicanonical simulation\index{multicanonical sampling}, which yields
a constant energy distribution $h_{\rm muca}(E)= \int_0^1dQ\,H_{\rm muca}(E,Q) \approx {\rm const.}$
In consequence, $H_{\rm muca}(E,Q)$ is useful for identifying the folding channels, 
independently of temperature. Restricting the canonical partition function at temperature $T$ to the ``microoverlap'' ensemble
with overlap $Q$, $Z(Q)= \int {\cal D}{\bf X}\, \delta(Q-Q({\bf X}))\,\exp\{-E({\bf X})/k_BT\}$,
where the integral is over all possible conformations ${\bf X}$, we define the overlap free energy 
as $F(Q)=-k_BT\ln Z(Q)$. 

Figures~\ref{fig:fandh}(a)--(c) show the thus obtained multicanonical histograms $H_{\rm muca}(E,Q)$ (left) 
and the overlap free-energy landscapes $F(Q)$ (right) at different temperatures for the three sequences listed in 
Table~\ref{tab:seqs}. The different branches of $H_{\rm muca}(E,Q)$ indicate the channels the heteropolymer 
can follow in the folding process towards the reference structure. 
The heteropolymers, whose sequences differ only by permutations, exhibit noticeable differences in the folding behavior 
towards the native conformations. The first interesting observation is that the minimalistic model used is
capable of revealing the different folding behaviors of the wild-type and permuted sequences. The second remarkable
result is that the angular overlap parameter $Q$ is a surprisingly manifest measure for the peptide macrostate.

From Fig.~\ref{fig:fandh}(a) we conclude that folding of sequence S1 exhibits a typical
two-state characteristics. Above the transition temperature, i.e., in the regime of denatured conformations D, 
conformations possess a random-coil-like overlap $Q\approx 0.7$, i.e, there is no significant 
similarity with the reference structure. Close to $T\approx 0.1$ the global minimum of the corresponding 
overlap free energy $F(Q)$ changes discontinuously towards larger $Q$ values, and 
at the transition state the denatured (D) and the folded macrostates (N) are equally probable. The existence of 
this pronounced transition state is a characteristic indication for first-order-like two-state folding. 
Decreasing the temperature further, the native-fold-like conformations ($Q>0.95$) dominate and fold smoothly 
towards the $Q=1$ reference structure, i.e., the lowest-energy conformation found for sequence S1.

The folding behavior of sequence S2 is significantly different, as Fig.~\ref{fig:fandh}(b) shows, and
is a typical example for a folding event through an intermediate macrostate. The main channel D 
bifurcates and a side channel I branches off continuously. For smaller energies (or lower temperatures), 
this branching is followed by the formation 
of a third channel N, which ends in the native fold. The characteristics of folding-through-intermediates
is also reflected by the free-energy landscapes. Starting at high temperatures in
the pseudophase of denatured conformations D with $Q\approx 0.76$,
the intermediary phase I with $Q \approx 0.9$ is reached close to the temperature $T\approx 0.05$.
Decreasing the temperature further below the native-folding
threshold close to $T=0.01$, the hydrophobic-core formation is finished and stable native-fold-like conformations with
$Q>0.97$ dominate in regime N.

\begin{figure}[t]
\centerline{\epsfxsize=8.8cm \epsfbox{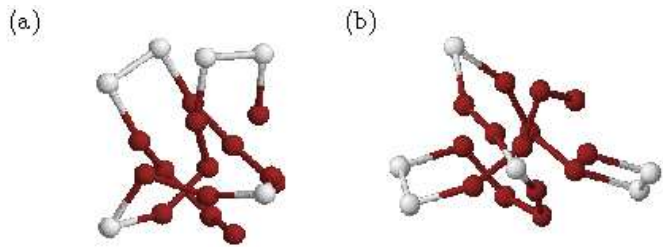}}
\caption{\label{fig:gemB} Lowest-energy conformations for sequence S3, considered as (a) reference 
structure ${\bf X}^{(0)}$
and (b) alternative metastable conformation, whose angular overlap with ${\bf X}^{(0)}$ is $Q\approx 0.75$.}
\end{figure}

The most extreme behavior of the three exemplified sequences is found for sequence S3, where
the main channel D does not decay in favor of a single native-fold channel. In fact, in Fig.~\ref{fig:fandh}(c) 
we observe both, {\em two} separate native-fold channels, M$_1$ and M$_2$, and a bifurcating main channel. 
Above the folding transition\index{transition!folding} ($T=0.2$), the typical 
sequence-independent denatured conformations in D ($Q\approx 0.77$) dominate. 
Annealing below the glass-transition\index{transition!glass} threshold, 
several channels form and coexist. The two most prominent channels (to which the lowest-energy
conformations belong that we found in the simulations) eventually lead for $T\approx 0.01$ to ensembles of states
M$_1$ with $Q>0.97$, which are similar to the reference structure shown in Fig.~\ref{fig:gemB}(a), and 
conformations M$_2$ with $Q\approx 0.75$. The lowest-energy conformation found in this regime is shown in 
Fig.~\ref{fig:gemB}(b). It is structurally different but energetically almost degenerate compared with the 
reference structure. 
It should also be noted that the lowest-energy main-channel conformations
have only slightly larger energies than the two native folds. Thus, the folding of this heteropolymer
is accompanied by a very complex, amorphous folding characteristics. In fact, the multiple-peaked distribution $H_{\rm muca}(E,Q)$
near minimum energies is a strong indication for metastability. A native fold in the natural sense
does not exist, the $Q=1$ conformation is only a reference structure but the folding towards this
structure is not distinguished as it is in the folding characteristics of sequences S1 and S2.
These results demonstrate that it is possible to find clear indications for three different 
folding characteristics known from real proteins by analyzing macrostates based on an angular overlap
parameter within a minimalistic heteropolymer frame. The physical objective is not only 
on establishing a quantitative one-to-one correspondence between model and real peptides (which, in general, 
is not in the focus
of minimalistic, effective models), but also on a more comprehensive, qualitative understanding of universal
aspects of protein folding. We find that for selected hydrophobic-polar
heteropolymer sequences 
characteristic folding behaviors such as two-state folding, folding through intermediates, 
and metastability can be observed which are qualitatively comparable with real folding events in nature.
Beyond the general interest in understanding complex aspects of protein folding, the preparation
of synthetic peptide macrostates in future applications, e.g., in the design of substrate- or pattern-selective 
polymers is strongly connected with the understanding of such 
conformational folding transitions\index{transition!conformational}\index{transition!folding}.
\section{Peptide aggregation}
\index{transition!aggregation}
\index{aggregation}
Another important, because biologically relevant, example for cooperative structure formation
processes is the aggregation of proteins. A prominent example, where this process has
disastrous effects, is the oligomerization of the A$\beta$ protein which is associated 
with Alzheimer's disease. 

\begin{figure}[t]
\centerline{\epsfxsize=10.0cm \epsfbox{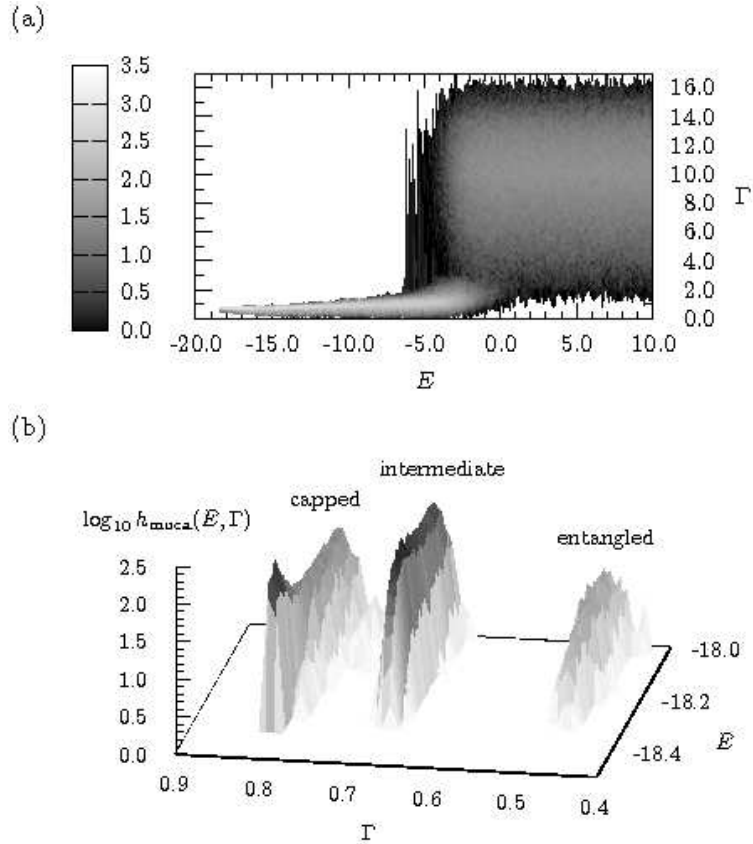}}
\caption{\label{fig:agghist} 
(a) Multicanonical histogram $\log_{10}h_{\rm muca}$ as
a function of energy $E$ and aggregation parameter $\Gamma$, (b) section of $\log_{10}h_{\rm muca}$
in the low-energy tail~\cite{jbj2}.}
\end{figure}

A mesoscopic model for the aggregation of multiple chains can simply be defined by 
assuming that the same type-dependent Lennard-Jones like potentials used in the 
single-chain form~(\ref{eq:ab}) describe also the inter-monomeric interaction, i.e., the
interaction among monomers of different chains~\cite{jbj1}. For the analysis of the
aggregation transition let us consider the example of a complex of two identical AB peptides
with sequence $AB_2AB_2ABAB_2AB$~\cite{jbj1,jbj2}. We suppose that the aggregation of the peptides should
be signaled by strong fluctuations of the relative distance of the centers of masses of the 
individual chains. Thus we define for systems consisting of $M$ peptides
\begin{equation}
\Gamma^2=\frac{1}{2M^2}\sum_{\mu,\nu=1}^{M}
\left({\bf r}_{\rm COM}^{(\mu)}-{\bf r}_{\rm COM}^{(\nu)}\right)^2,
\end{equation}
where ${\bf r}_{\rm COM}^{(\mu)}$ is the center of mass of the $\mu$th chain (in our example
$M=2$). Actually, a multicanonical computer simulation reveals very clear indications
for a single conformational transition, the aggregation transition~\cite{jbj1,jbj2}. This means that the 
peptide-peptide aggregation and the folding into a compact peptide complex are not separate
transitions (at least in this example). This is illustrated in Fig.~\ref{fig:agghist}(a),
where the color-coded multicanonical histogram as a function of energy $E$ and the aggregation
parameter $\Gamma$ is shown. Qualitatively, two separate main branches 
(which are ``channels'' in the corresponding free-energy landscape) are apparent, between which a noticeable
transition occurs. In the vicinity of the energy $E_{\rm sep}\approx -3.15$, both channels overlap, i.e., 
the associated macrostates coexist. Since $\Gamma$ is an effective measure for the spatial distance between
the two peptides, it is obvious that conformations with separated or fragmented peptides belong to 
the dominating channel in the regime of high energies and large $\Gamma$ values, whereas the aggregates are
accumulated in the narrow low-energy and small-$\Gamma$ channel. Thus, the main observation from the 
multicanonical, comprising point of view is that the aggregation transition is a phase separation process
which already appears, even for this small system, in a surprisingly clear fashion. 

\begin{figure}
\centerline{\epsfxsize = 7.5cm \epsfbox{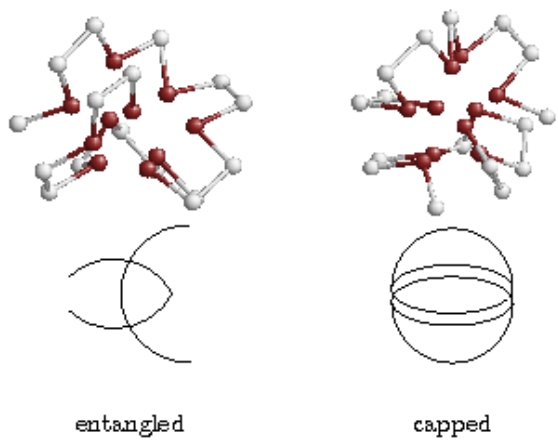}}
\caption{\label{fig:aggconfs} Representatives and schematic characteristics of entangled and 
spherically-capped conformations dominating the lowest-energy branches in the multicanonical
histogram shown in Fig.~\ref{fig:agghist}(b). Dark spheres correspond to hydrophobic ($A$), 
light ones to polar ($B$) residues.}
\end{figure}

The high precision
of the multicanonical method allows us even to see further details in the lowest-energy aggregation regime,
which is usually a notoriously difficult sampling problem. Fig.~\ref{fig:agghist}(b) shows that the tight
aggregation channel splits into three separate, almost degenerate subchannels at lowest energies. 
From the analysis
of the conformations in this region, one finds that representative conformations with smallest $\Gamma$ values,
$\Gamma\approx 0.45$, are typically entangled, while those with $\Gamma\approx 0.8$ have a spherically-capped
shape. Examples are shown in Fig.~\ref{fig:aggconfs}. The also highly compact conformations belonging to the 
intermediate subphase do not exhibit such 
characteristic features and are rather globules without noticeable internal symmetries. In all cases, 
the aggregates contain a single compact core of hydrophobic residues. This also confirms that 
the aggregation is not a simple
docking process of two prefolded peptides, but a complex cooperative folding-binding process. The 
general aggregation behavior is similar also for larger systems of more peptides with the
same sequence~\cite{jbj2}.
\section{Summary}
\label{sect:sum}
For the qualitative analysis of phase transitions, it is often sufficient to 
perform statistical studies of simplified effective models, where the natural
complexity of the realistic system is broken down to the essential, irreducible
level of cooperative behavior. The probably most famous example is the Ising model of ferromagnetism.
In this model, a local short-range spin-spin interaction -- which in essence 
is a consequence of the quantum mechanical exchange mechanism between
magnetic moments -- triggers in two and more dimensions a nontrivial second-order
phase transition between the ordered ferromagnetic macrostate and the disordered,
random paramagnetic phase. The generalization of the description of phase transitions
is highly successfully achieved within the framework of Ginzburg-Landau theories,
which are not only restricted to transitions of second order, but also allow
investigations of symmetry breaking typically forcing first-order phase transitions.
In any case, the idea is to introduce collective coordinates, or, more
specific, order parameters that allow for a unique identification of the 
actual macrostate of the system.

The characterization of conformational (structural) transitions\index{transition!conformational} during the
folding process of proteins is more involved as no general theory of phase transitions
for finite systems is available. In fact, the finiteness of the amino acid sequence 
length contradicts the demand of a thermodynamic
limit, which is the essential condition for thermodynamic phase transitions to occur. Nonetheless,
there is hope that following a similar strategy as in the theory for phase transitions,
a classification of characteristic tertiary folding transitions\index{transition!folding} (e.g., single-exponential
folding, two-state folding, folding through weakly stable intermediary states, 
metastability) is possible. If so, then it should be possible to construct simple models at a raw,
coarse-grained level that allow firstly the introduction of unique conformational (``order'')
parameters and secondly to qualitatively reproduce the known folding characteristics
of classes of proteins. 

As the Ising model will not be an adequate model
for precise questions regarding a {\em specific} ferromagnet, it is also not
expected that a simple, coarse-grained model will reveal the folding behavior of
a {\em specific} protein. This means, for explaining the folding characteristics of a 
{\em specific} protein, doubtlessly a microscopic all-atom model incorporating interactions 
acting over all length and energy scales is required. 

In this lecture, we have demonstrated, however, that results obtained from simple lattice and
off-lattice heteropolymer models are indeed capable of revealing characteristic features 
of proteins (stability of designing sequences, designable conformations) and 
protein folding\index{protein folding} 
(folding channels, free-energy landscapes). As far as important qualitative features
of peptides and proteins on intermediate length scales are concerned, such models are thus of
comparable significance as the more detailed atomic descriptions. 
\section*{Acknowledgments}
We are grateful to Peter Grassberger and Hsiao-Ping Hsu for detailed informations 
about PERM and its improved variants. We are also indebted to Anders Irb\"ack, Sandipan
Mohanty, and Simon Mitternacht for
discussions on coarse-grained and simplified variants of microscopic protein models,
and to Bernd A.\ Berg, Ulrich H.\ E.\ Hansmann, Yuko Okamoto, Tar{\i}k \c{C}elik, and
G\"{o}khan G\"{o}ko\u{g}lu for discussions on general aspects of protein folding.
We thank Handan Ark{\i}n,
Reinhard Schiemann, Thomas Vogel, Stefan Schnabel, Anna Kallias, Jakob Schluttig,
and Christoph Junghans for cooperation in studies of coarse-grained lattice and off-lattice 
heteropolymer models. The investigations of hybrid interfaces were inspired by
experiments of Karsten Goede and Marius Grundmann, which we would like to thank for
close collaboration.
This work is partially supported by the DFG (German Science Foundation) grant
under contract No.\ JA 483/24-1 and by the DAAD-STINT Personnel Exchange Programme
with Sweden. M.B.\ thanks the DFG and the Wenner-Gren Foundation for research fellowships.
Support by the JUMP supercomputer time grant No.\ hlz11
of the John von Neumann Institute for Computing (NIC), Forschungszentrum
J\"ulich, is also gratefully acknowledged.
%
%
%

%
%

\begin{thebibliography}{99}
\bibitem{rem1}
Two sequences are only distinguished, if they are not
symmetric under reversal of their residues. For a chain with length $N=4$, for example, there are 
10 relevant sequences instead of $2^4=16$.
\end{thebibliography}
\end{document}